\documentclass[intlimits,twoside,a4paper]{article}

\usepackage{amsmath,amssymb}
\usepackage{graphicx}

\usepackage[T2A]{fontenc}
\usepackage[cp1251]{inputenc}

\usepackage[eqsecnum]{cmpj2}

\issue{2017}{20}{1}{13802}
\doinumber{10.5488/CMP.20.13802}
\title[Shape characteristics]{Shape characteristics of the aggregates formed by amphiphilic stars in water: dissipative particle dynamics study\thanks{This paper is written in honor of Yurij Holovatch on the occasion of his 60th birthday.}}
\author[O.Y. Kalyuzhnyi, J.M. Ilnytskyi, C. von Ferber]{O.Y. Kalyuzhnyi\refaddr{label1,label4},
        J.M. Ilnytskyi\refaddr{label1,label4}, C. von Ferber\refaddr{label2,label3,label4}}
\addresses{
\addr{label1} Institute for Condensed Matter Physics
of the National Academy of Sciences of Ukraine,\\
1 Svientsitskii St., 79011 Lviv, Ukraine
\addr{label2} Applied Mathematics Research Centre, Coventry University, Coventry, CV1 5FB, United Kingdom
\addr{label3} Heinrich-Heine Universit\"at D\"usseldorf, D-40225 D\"usseldorf, Germany
\addr{label4} Doctoral College for the Statistical Physics of Complex Systems, Leipzig-Lorraine-Lviv-Coventry
              $({\mathbb L}^4)$,\\ D-04009 Leipzig, Germany
}

\newcommand{\vvec}[1]{\mathbf{#1}}
\newcommand{\idx}[1]{_{\mathrm{#1}}}

\DeclareMathOperator{\Tr}{Tr}

\authorcopyright{O.Y. Kalyuzhnyi, J.M. Ilnytskyi, C. von Ferber, 2017}
\date{Received January 20, 2017, in final form February 16, 2017}

\begin{document}

\maketitle

\begin{abstract}
We study the effect of the molecular architecture of amphiphilic star polymers on
the shape of aggregates they form in water. Both solute and solvent are considered
at a coarse-grained level by means of dissipative particle dynamics simulations.
Four different molecular architectures are considered: the miktoarm star, two different
diblock stars and a group of  linear diblock copolymers, all of the same composition and
molecular weight. Aggregation is started from a closely packed bunch of $N_{\text a}$ molecules
immersed into water. In most cases, a single aggregate is observed as a result of
equilibration, and its shape characteristics are studied depending on the aggregation number
$N_{\text a}$. Four types of aggregate shape are observed:
spherical, rod-like and disc-like micelle and a spherical vesicle. We estimate ``phase boundaries''
between these shapes depending on the molecular architecture. Sharp transitions between
aspherical micelle and a vesicle are found in most cases. The pretransition region shows
large amplitude oscillations of the shape characteristics with the oscillation frequency
strongly dependent on the molecular architecture.
\pacs 82.70.Uv, 78.67.Ve, 61.20.Ja
\end{abstract}

%==============================
\section{\label{I}Introduction}
%==============================

Star polymers represent one of the simplest branched polymeric architectures
utilising several arms in the form of linear chains linked to a central core. Branching imposes
intramolecular constraints on the monomers and modifies mechanical,
viscoelastic, and solution properties of such molecules in bulk and in a solution. A relatively recent review of the
synthesis, properties and application of star polymers is given in
\cite{Hadjichristidis2012}.

Recently, one observes a growing interest towards amphiphilic and polyphilic
star polymers. There are two main types of such star polymers, namely, the
miktoarm and the diblock stars. In the miktoarm star, all the arms are of a homopolymer type
but their properties differ from arm to arm. In the diblock star, each arm is
a linear diblock copolymer itself. Both cases generate interest in both synthetic
methodologies and in the self-assembly of such molecules in solution
\cite{Khanna2010,Moughton2012,Park2014}. In particular, depending on the details
of the molecular architecture, a number of morphologies are observed, such as:
Archimedean tiling patterns and cylindrical microdomains for miktoarm star copolymers
as well as asymmetric lamellar microdomains for diblock stars, which have not
been reported for linear block copolymers, for more details see \cite{Park2014}.

Applications of star polymers range from nano- and micropatterning
\cite{Hadjichristidis2012,Park2014} in gel and bulk state to drug delivery via
their micellisation in solution (see, e.g. \cite{Kozlovskaya2016}).
In the latter case, both {\it the size and the shape of a micelle} plays an important
role from the requirements of low toxicity and efficient transport of the drug.
For example, it has been reported that smaller nanoparticles ($\sim 25$~nm)
travel through the lymphatic more readily than the larger particles ($\sim 100$~nm)
and accumulate in lymph node resident dendritic cells \cite{Reddy2007,Zolnik2010}.
Besides that, the shape of the particles also affect their bio-application:
e.g., worm-shaped filamentous micelles show less phagocytosis as compared to spheres,
rod-like and red blood cell discoidal particles boasted longer blood circulation
time than spheres, which reduced their clearance from the bloodstream
(see \cite{Kozlovskaya2016} and references therein).

Besides experimental studies, the aggregation of polyphilic star polymers has been a
subject of theoretical investigations \cite{Ma2007} as well as computer simulations
using various approaches. In particular, Monte Carlo simulations \cite{Binder2008,Patti2010}
performed on a lattice revealed a spontaneous formation of roughly spherical aggregates of
diblock stars in a selective solvent. Significant changes in the micellar properties,
such as the critical micellar concentration and aggregation number are reported upon
changing the fraction of the solvophobic part of the diblock while keeping the total
arm length constant.

Coarse-grained approaches, such as the dissipative particle dynamics (DPD)
\cite{Hoogerbrugge1992,Espanol1995}, make it possible to consider both a variety
of molecular architectures and chemical compatibility of the particular groups,
combined with the computational efficiency. As a result, one can study microphase separation
driven effects in amphi- and polyphilic molecules, self-assembly, adsorption, etc.
(see, e.g. \cite{Ilnytskyi2008,Ilnytskyi2011,Ilnytskyi2013,Ilnytskyi2016}).
With respect to the aggregation of the polyphilic star molecules, there are a number of
studies of polymerosomes \cite{Ortiz2005}, ABC star block copolymers
\cite{Cui2006,Xia2006a,Xia2006b}, as well as related systems, e.g., a mixture of diblock
and homopolymers \cite{Zhao2009}. A number of micellar shapes have been observed, such as
spherical micelles, worm-like micelles, hamburgers and others.

Therefore, coarse-grained DPD simulations can be considered as a useful instrument to
study the size and the shape of aggregates formed of amphi- and polyphilic molecules of
star architecture and, therefore, are capable of sheding more light on their applicability for
micelle-based drug delivery systems. The aim of this work is to study the effect of the
molecular architecture of amphiphilic star polymers on the type of aggregates they
form in water. We consider three types of stars: the miktoarm star and two different diblock
stars. These are compared against the equivalent set of linear diblock copolymers.
The molecular weight and the fraction of hydrophilic monomers are chosen to be the same in all cases.
The outline of the study is as follows. The simulation approach and the properties of interest are
described in section~\ref{II}, the results are presented in section~\ref{III} followed by conclusions in
section~\ref{IV}.

%=================================================================
\section{\label{II}Simulation approach and properties of interest}
%=================================================================

We employ the mesoscopic method of DPD \cite{Hoogerbrugge1992,Espanol1995}
simulations, which describes polymer molecules at the level of coarse-grained beads
each representing a fragment of the chain. The same applies to the water
solvent, in which case a single solvent bead is assumed to contain
several molecules of water. All beads representing both a polymer and
water are spheres of the same diameter which provides the length-scale
of the problem, whereas the energy scale is assumed to be $\epsilon^*=k_{\text B}T=1$,
where $k_{\text B}$ is the Boltzmann constant, $T$ is the temperature, and time
is expressed in $t^*=1$ \cite{Groot1997}. The monomers are connected \textit{via}
harmonic springs, which results in a force
\begin{equation}\label{FB}
  \vvec{F}^{\text B}_{ij} = -k\vvec{x}_{ij}\,,
\end{equation}
where $\vvec{x}_i$ are the coordinates of $i$th bead, $\vvec{x}_{ij}=\vvec{x}_i-\vvec{x}_j$
and $k$ is the spring constant. The non-bonded forces contain three contributions
\begin{equation}
  \vvec{F}_{ij} = \vvec{F}^{\,\mathrm{C}}_{ij} + \vvec{F}^{\,\mathrm{D}}_{ij}
  + \vvec{F}^{\,\mathrm{R}}_{ij}\,,
\end{equation}
where $\vvec{F}^{\,\mathrm{C}}_{ij}$ is the conservative force
resulting from the repulsion between $i$th and $j$th soft beads,
$\vvec{F}^{\,\mathrm{D}}_{ij}$ is the dissipative force that occurs
due to the friction between soft beads, and random force
$\vvec{F}^{\,\mathrm{R}}_{ij}$ that works in pair with a dissipative
force to thermostat the system. The expressions for all these three
contributions are given below \cite{Groot1997}
\begin{equation}\label{FC}
  \vvec{F}^{\,\mathrm{C}}_{ij} =
     \left\{
     \begin{array}{ll}
        a(1-x_{ij})\displaystyle\frac{\vvec{x}_{ij}}{x_{ij}}\,, & \quad x_{ij}<1,\\
        0,                       &\quad x_{ij}\geqslant 1,
     \end{array}
     \right.
\end{equation}
\begin{equation}\label{FD}
  \vvec{F}^{\,\mathrm{D}}_{ij} = -\gamma
  w^{\,\mathrm{D}}(x_{ij})(\vvec{x}_{ij}\cdot\vvec{v}_{ij})\frac{\vvec{x}_{ij}}{x^2_{ij}}\,,
\end{equation}
\begin{equation}\label{FR}
  \vvec{F}^{\,\mathrm{R}}_{ij} = \sigma
  w^{\,\mathrm{R}}(x_{ij})\theta_{ij}\Delta t^{-1/2}\frac{\vvec{x}_{ij}}{x_{ij}}\,,
\end{equation}
where $x_{ij}=|\vvec{x}_{ij}|$,
$\vvec{v}_{ij}=\vvec{v}_{i}-\vvec{v}_{j}$, $\vvec{v}_{i}$ is the
velocity of $i$th bead, $a$ is the amplitude for the conservative
repulsive force. The dissipative force has an amplitude $\gamma$ and
decays with the distance according to the weight function
$w^{\,\mathrm{D}}(x_{ij})$. The amplitude for the random force is
$\sigma$ and the respective weight function is
$w^{\,\mathrm{R}}(x_{ij})$. $\theta_{ij}$ is the Gaussian random
variable and $\Delta t$ is the time-step of the simulations. As was
shown by Espa\~{n}ol and Warren \cite{Espanol1995}, to satisfy the
detailed balance requirement, the amplitudes and weight functions
for the dissipative and random forces should be interrelated:
$\sigma^2=2\gamma$ and
$w^{\,\mathrm{D}}(x_{ij})=[w^{\,\mathrm{R}}(x_{ij})]^2$.

We will denote the beads representing water as of type A.
The compressibility of such coarse-grained solvent at a number density of
beads equal to $\rho=3$ matches that for water at normal condition
if the repulsion parameter $a$ in equation~(\ref{FC}) is chosen equal
to $a_{\text{AA}}=25$ \cite{Groot1997}.
Similarly, the level of hydrophilicity of a polymer fragment composed of beads
of type P can be controlled by the value $a_{\text{PA}}$ for the repulsion interaction
between beads P and A, where the difference $a_{\text{PA}}-a_{\text{AA}}$ is related to the
Flory-Huggins parameter \cite{Groot1998}. In our study, we assume the hydrophilic
fragments of a polymer chain to be composed of the beads A, the same as for
water, whereas the hydrophobic fragments --- of the beads B, characterised by the
repulsion parameter $a_{\text{AB}}=40$, the value already used in a number of studies
\cite{Ilnytskyi2007,Ilnytskyi2011,Ilnytskyi2013}. The other parameters are
as follows: $\gamma=6.75$, $\sigma=\sqrt{2\gamma}=3.67$ and the time-step
$\Delta t=0.04$. Another parametrisation of the repulsive potential in DPD studies
aimed at polymerosomes can be found in \cite{Ortiz2005}.

We use the simulation box with the periodic boundary conditions. Its
linear dimension $L$ is chosen at least twice the linear dimension of the largest
aggregate formed out of $N\idx{mol}$ molecules each containing $n$ beads.
This requirement appears from the need to restore the integrity of an aggregate
if it is split across one or several walls of the simulation box.
A rough estimate for $L$ can be achieved by assuming the
formation of a single spherical aggregate with the diameter $D_{\text a}$.
As far as each bead on average occupies the volume of $1/\rho$, the volume of an
aggregate is $nN\idx{mol}/\rho=\pi D_{\text a}^3/6$, hence
$D_{\text a}\sim(6nN\idx{mol}/\pi\rho)^{1/3}$ and, therefore, the condition for
the linear dimension of the simulation box reads $L > 2D_{\text a}$. For highly aspherical
aggregates, the value of $L$ should be accordingly increased.

To identify the existing aggregates at each time instance, we use the following algorithm:
\begin{enumerate}
\renewcommand{\labelenumi}{(\arabic{enumi})}
 \item for each $(i\idx{mol}, j\idx{mol})$ pair of molecules, find the number $n_{\text c}$ of close bead-bead
       intermolecular contacts evaluated for hydrophobic beads (type B), and if $n_{\text c}>n\idx{min}$, then
       register the link between the molecules $i\idx{mol}$ and $j\idx{mol}$;
 \item using the linking list for molecular pairs, build neighbour lists for linked molecules;
 \item apply a ``snowball'' cluster identification algorithm, in which the neighboring molecules are added
       to the existing ones within each aggregate until no neighbours remain;
 \item final information contains: the lists of molecules that belong to a particular $k$th aggregate, and
       the index array to determine the aggregate to which any molecule belongs.
\end{enumerate}
A close bead-bead contact is registered when the distance $x_{ij}$ between them is less than $1.5$,
and the threshold number of close contacts $n\idx{min}$ is chosen to be equal to $10$.

After the aggregates have been identified, the lists of molecules for each of it is used to
rejoin the aggregates split by the periodic boundary conditions. Then,
we proceed to the analysis of their size and shape properties. The radius of gyration and
the shape characteristics of each aggregate are derived from the components of the gyration
tensor $\vvec{Q}$ defined as in \cite{Solc1971a,Solc1971b}:
\begin{equation}\label{Q_def}
Q_{\alpha\beta} = \frac{1}{N}
\sum_{n=1}^{N}\left(x^{\alpha}_{n}-X^{\alpha}\right)\left(x^{\beta}_{n}-X^{\beta}\right),
\qquad \alpha,\beta=1,2,3.
\end{equation}
Here, $N$ is the total number of beads in an aggregate, $x_n^{\alpha}$
denotes the coordinates of $n$th bead: $\vvec{x}_n=(x_n^1,x_n^2,x_n^3)$,
and $X^{\alpha} = \frac{1}{N}\sum_{n=1}^N x_{n}^{\alpha}$ are the coordinates
of the center of mass for the aggregate. Its eigenvectors define the axes of a
local frame of the chain and the mass distribution of the latter along
each axis is given by the respective eigenvalue $\lambda_i$,
$i=1,2,3$, respectively. The trace of $\vvec{Q}$ is an invariant
with respect to rotations and is equal to an instantaneous squared
gyration radius of the chain
\begin{equation}\label{Rg_def}
R_{\text g}^2 = \Tr\vvec{Q} = 3\bar{\lambda}.
\end{equation}
Here, the average over three eigenvalues, $\bar{\lambda}$, is
introduced to simplify the following expressions.
The hydrodynamic radius $R_{\text h}$ is found according to the following expression
\begin{equation}\label{Rh_def}
R^{-1}_{\text h} = \Big\langle \frac{1}{r_{ij}}\Big\rangle,
\end{equation}
where the averaging is performed over all pairs of beads that belong to the aggregate.

The shape properties of aggregates are characterised by the asphericity $A$
(sometimes also referred to as ``the relative shape anisotropy'') and prolateness $S$
\cite{Aronovitz1986,Zifferer1999a,Zifferer1999b,Blavatska2011}, as well as the shape
descriptor $B$ introduced in this study:
\begin{equation}\label{ASB_def}
A = \frac{1}{6}\frac{\sum_{i=1}^{3}(\lambda_{i}-\bar{\lambda})^{2}}{\bar{\lambda}^{2}}\,,\qquad
S = \frac{\prod_{i=1}^{3}(\lambda_{i}-\bar{\lambda})}{\bar{\lambda}^{3}}\,,\qquad
B = - \frac{\lambda_{2}-\bar{\lambda}}{\bar{\lambda}}\,.
\end{equation}
For a spherical shape: $A=S=B=0$, whereas for the ideal rod ($\lambda_1=\lambda$, $\lambda_{2,3}=0$):
$A=1$, $S=2$, $B=1$ and for the ideal disc ($\lambda_{1,2}=\lambda$, $\lambda_3=0$):
$A=1/4$, $S=-1/4$, $B=-1$. The convenience of the use of the shape descriptor $B$ is
that its range is symmetric spanning from $-1$ to $1$ and it reverses its sign when the
shape changes from the disc-like to the rod-like.

At each time instant $t$, the largest aggregate, or ``the giant component'' in percolation terminology,
is identified and the set of its characteristics, $R^2_{\text g}(t)$, $R^{-1}_{\text h}(t)$, $A(t)$,
$S(t)$ and $B(t)$ are saved. Based on their behaviour with time $t$, we estimate the time
$t\idx{equ}$ needed for the aggregate to forget its initial state and to start displaying stationary
oscillatory behaviour in terms of its shape properties. The conservative estimate reads
$t^*\idx{ini}\sim 8000$ ($200\cdot 10^3$ DPD steps). Then, the average values $\langle R^2_{\text g}\rangle$,
$\langle R^{-1}_{\text h}\rangle$, $\langle A\rangle$, $\langle S\rangle$ and $\langle B\rangle$ are
evaluated by simple averaging of the instant values over the time interval $t\idx{equ}<t<t\idx{fin}$,
where $t^*\idx{fin}\sim 10 t^*\idx{equ}$ is the total duration of each run.
One should note that the other type of averaging is also possible, where
the nominators and denominators in equations~(\ref{ASB_def}) are averaged separately
\cite{Aronovitz1986,Jagodzinski1992,Zifferer1999a,Zifferer1999b,Blavatska2011,Kalyuzhnyi2016}.
Besides the simple averages, we also consider the histograms of probability distributions for the
shape characteristics, where the $p(B)$, for the shape descriptor $B$, is found to be the most informative.

%==============================================================
\section{Aggregates and their properties}\label{III}
%==============================================================
\begin{figure}[!b]
\begin{picture}(0,75)
\put(15,10){\includegraphics[width=2.3cm]{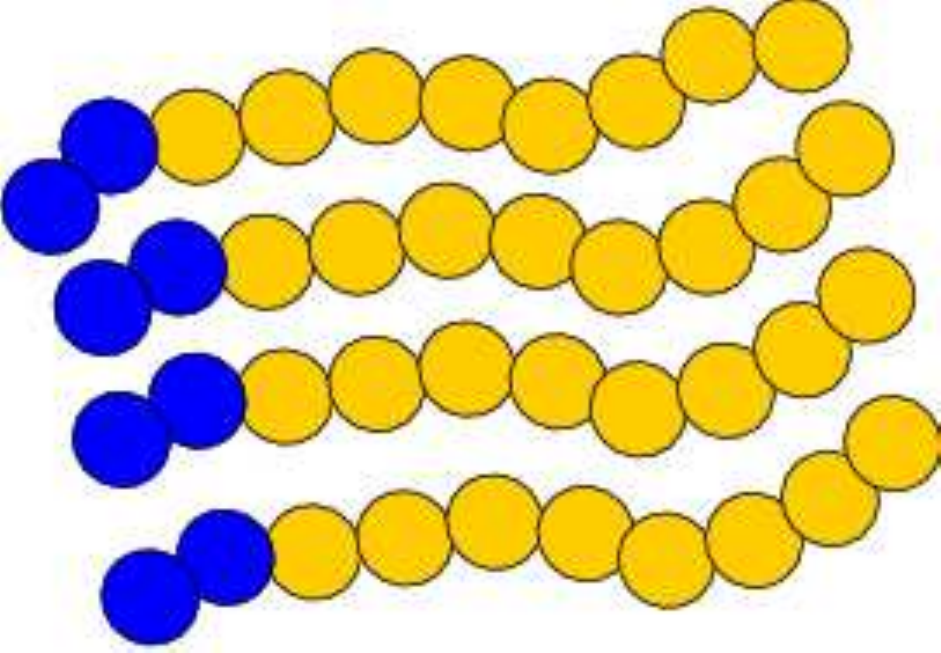}}
\put(95,10){\includegraphics[width=2cm]{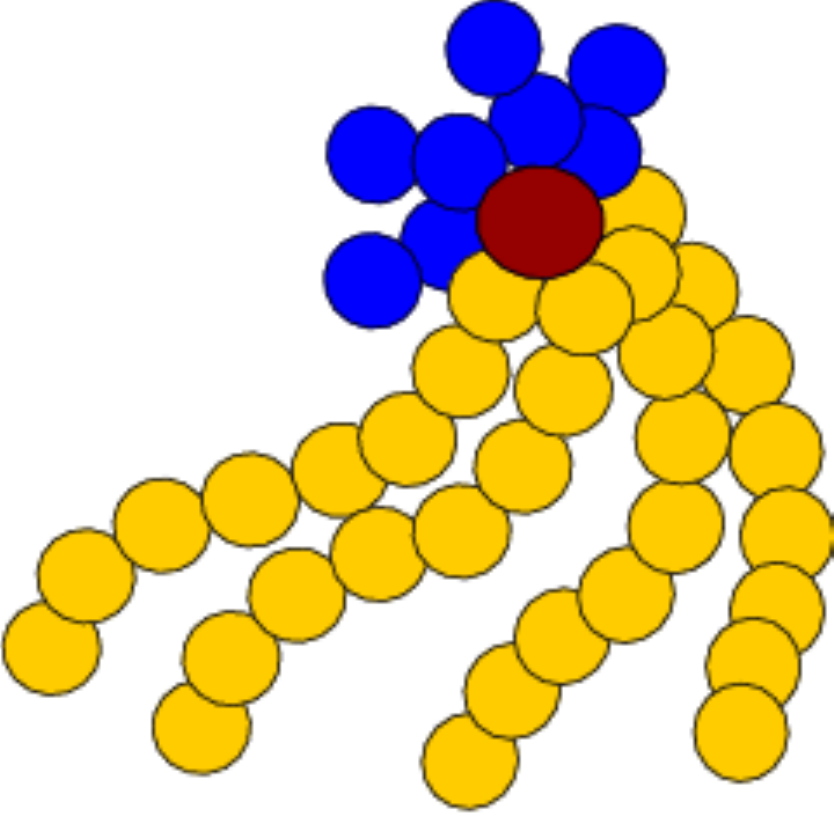}}
\put(170,0){\includegraphics[width=4cm]{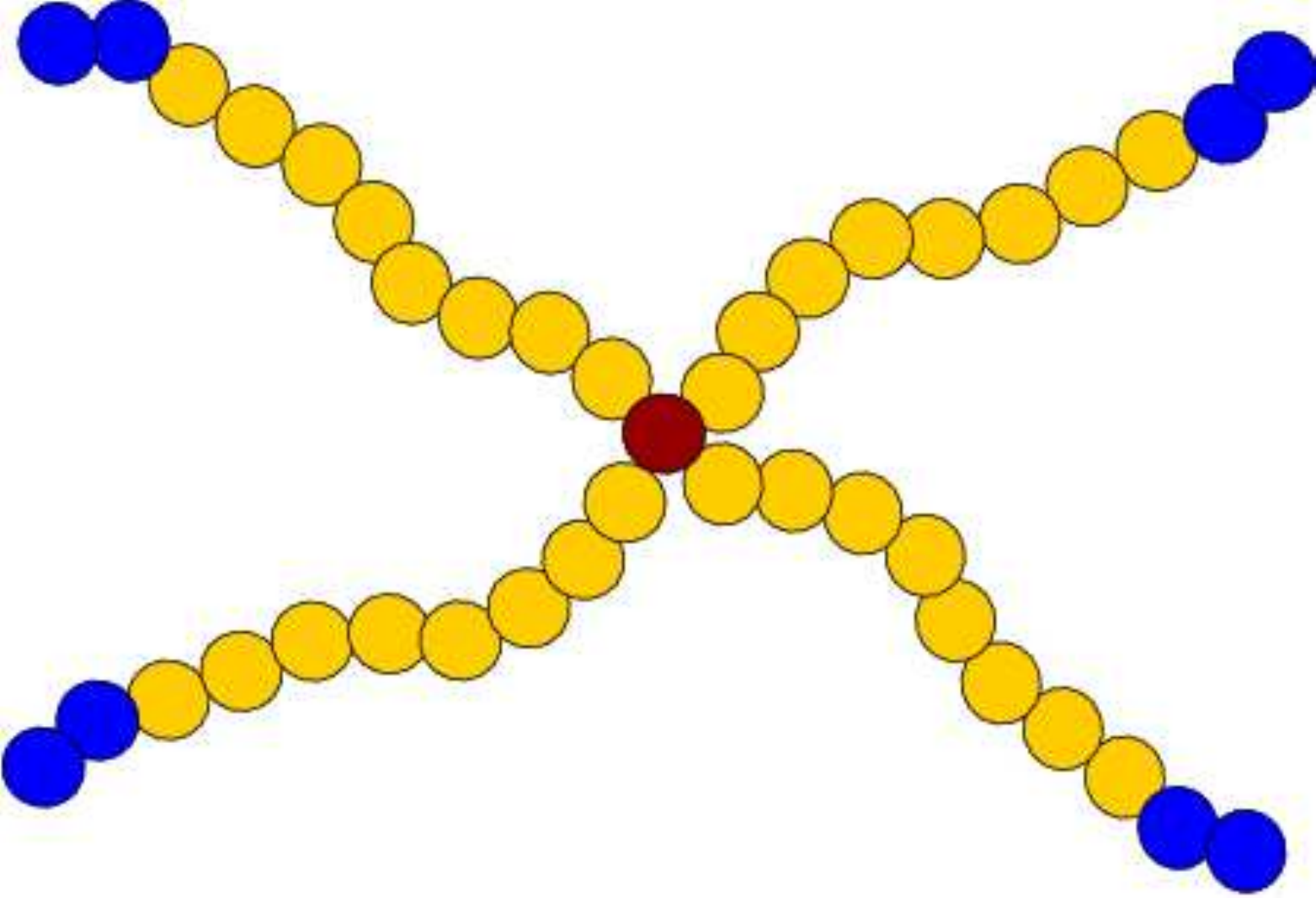}}
\put(300,0){\includegraphics[width=3.7cm]{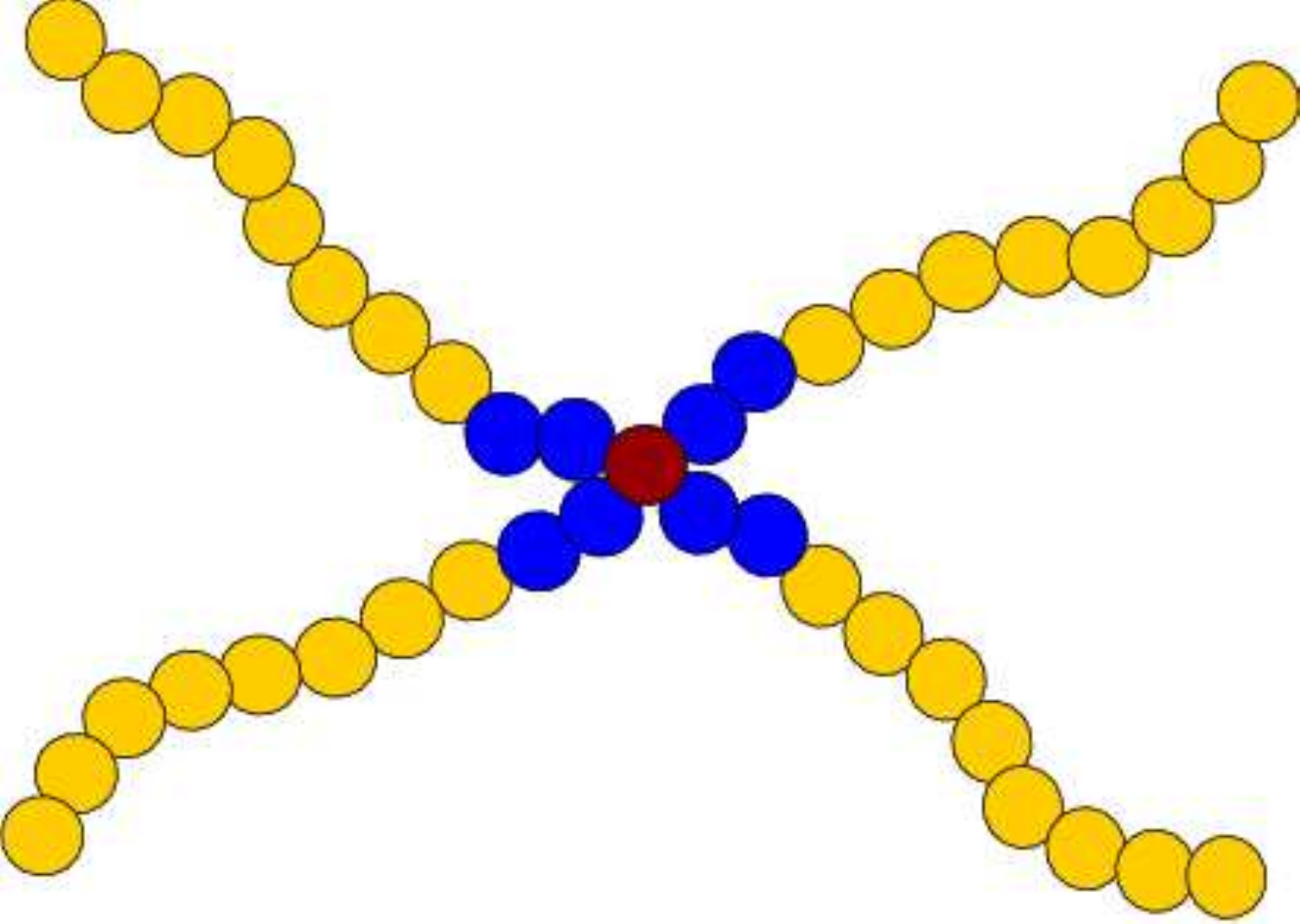}}
\put(40,75){(a)}
\put(120,75){(b)}
\put(222,75){(c)}
\put(345,75){(d)}
\end{picture}
\caption{\label{mol_arch} (Color online) Molecular architectures used in this study: (a)~linear diblock copolymers, (b)~miktoarm star-polymer, (c)~diblock 1~star-copolymer and (d)~diblock 2~star-copolymer. Colour coding: hydrophilic
beads (type A) are shown in blue, hydrophobic (type B) --- in yellow, central
bead is shown in red.}
\end{figure}
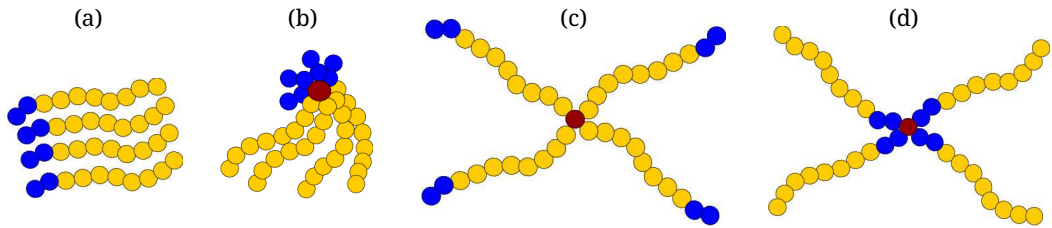

We consider four types of molecular architectures, all shown in figure~\ref{mol_arch}.
All of them can be interpreted as four linear diblock copolymers, each of eight hydrophobic beads
(type B) and of two hydrophilic beads (type A) bonded in a different way.
In particular, architecture (a) is just a set of four non-bonded diblocks and serves as some
kind of a reference system. Architecture (b) represents the bonding of four diblocks in the form
of an asymmetric miktoarm star of eight arms: four hydrophilic and four hydrophobic.
The (c) and (d) architectures are of the diblock star type with the arms being bonded
by either their hydrophobic ends, as in case (c), or by their hydrophilic ends, case (d).
Molecular mass in all four cases is practically the same, save for an additional central bead
in the cases (b)--(d), as compared to the case (a).

\begin{figure}[!t]
\begin{picture}(0,130)
\put(10,140){\includegraphics[width=5cm,angle=270]{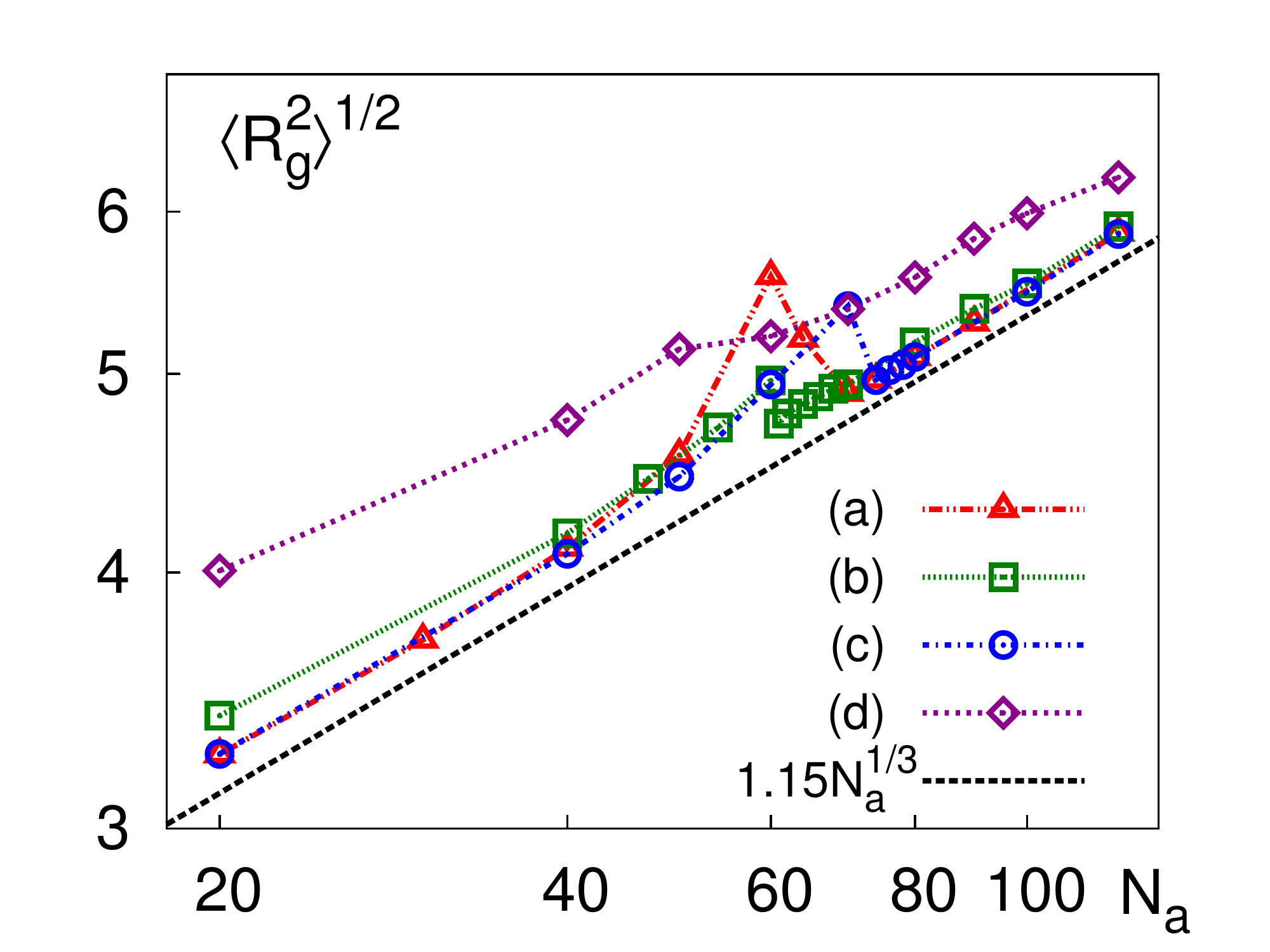}}
\put(210,140){\includegraphics[width=5cm,angle=270]{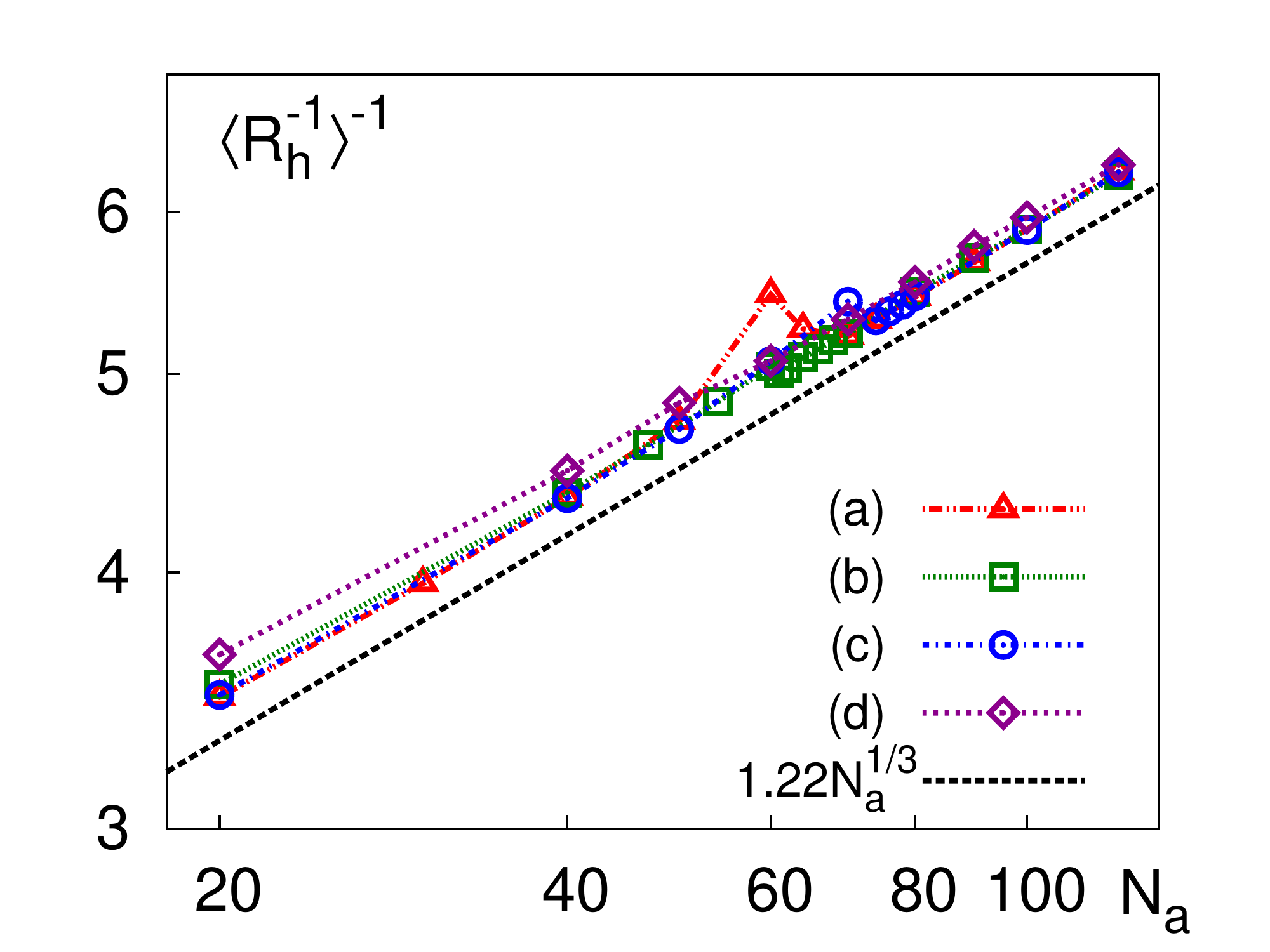}}
\put(100,120){(i)}
\put(300,120){(ii)}
\end{picture}
\caption{\label{Rg_Rh} (Color online) Comparison of the estimates for the average radius or gyration
[frame (i)] and hydrodynamic radius [frame (ii)] as functions of the aggregation number
$N_{\text a}$ performed for the molecular architectures listed in figure~\ref{mol_arch}. The collapse-like
scaling law, $N_{\text a}^{1/3}$, is provided in both cases as a guide for the eye.}
\end{figure}
The aggregation transition in a solution of polymers is of much interest and is
the subject of numerous recent works \cite{Patti2010,Zierenberg2014,Zierenberg2015,Zierenberg2016}.
In this study, we consider the setup that mimics a drop of a highly concentrated polymer solution immersed
into water. The drop is modelled via a bunch of $N\idx{mol}$ molecules closely packed
inside a central part of the simulation box. Namely, a central bead of each star is positioned randomly
with all its coordinates falling into the interval of $[L/3;\,2L/3]$. Each arm is then generated
as a random walk. This algorithm results in inevitable overlaps of beads in the initial configuration.
The overlaps, however, do not produce forces of large magnitude, as it would occur in atomic simulations,
due to the soft nature of the interaction potential (\ref{FC}). In the course of simulation, this bunch of molecules
looses the memory concerning its initial state during the time $t\idx{equ}$. After that it starts to display
stationary oscillations in its shape properties. The values for $t\idx{ini}$ and for the simulation duration
$t\idx{fin}$ have already been provided in section~\ref{II}.

In most cases being studied, the initial setup equilibrates into the single aggregate state.
This, however, might be the result of a finite system size and also a finite simulation time.
In a case of a single aggregate, the aggregation number $N_{\text a}$ coincides with the total number
of star molecules $N\idx{mol}$ in the solution. The increase of the latter results in
a growth of the average aggregate size characterised by the estimates for the gyration and
hydrodynamic radii, as shown in frames (i) and (ii) of figure~\ref{Rg_Rh}, respectively. We find
the respective data points obtained for the molecular architectures (a)--(c) to be very close for both
size properties being considered, especially at larger $N_{\text a}$. Therefore, the difference in
intramolecular binding in these cases has a minor impact on the average aggregate size.
The aggregate size is dominated by a collapse of the hydrophobic part of stars characterised
by the scaling laws $\langle R_{\text g}^2\rangle^{1/2}$, $\langle R_{\text h}^{-1}\rangle^{-1}\sim (n\idx{h}N_{\text a})^{1/3}$, where
$n_{\text h}$ is the number of hydrophobic beads in each star. Data points obtained for the case (d) are
found to be higher than the respective values for the cases (a)--(c), especially in frame (i).
This deviation, as well as local deviations from the scaling law observed at specific values of $N_{\text a}$
in cases (a)--(c) are related to the peculiarities of the shape oscillations discussed below.

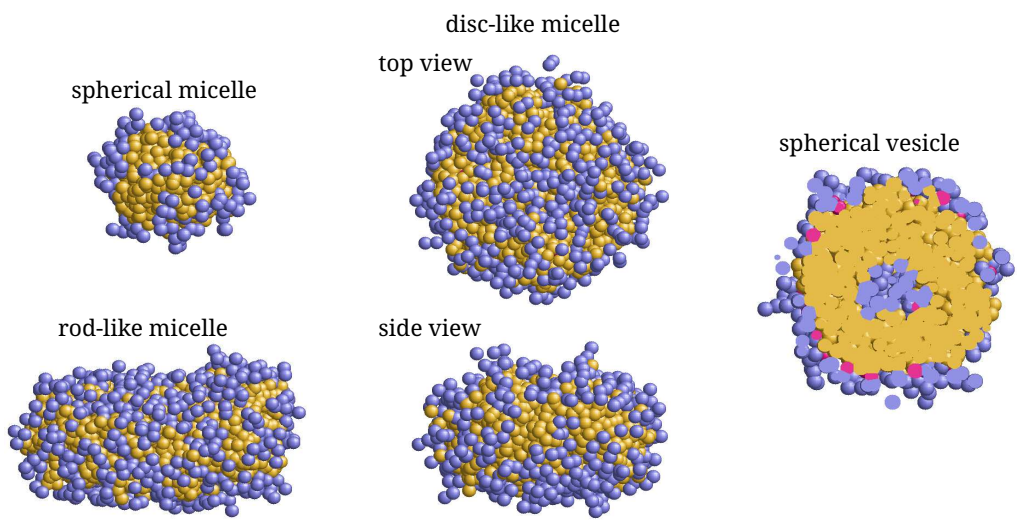
\begin{figure}[!t]
\begin{picture}(0,190)
\put(50,100){\includegraphics[width=2.3cm]{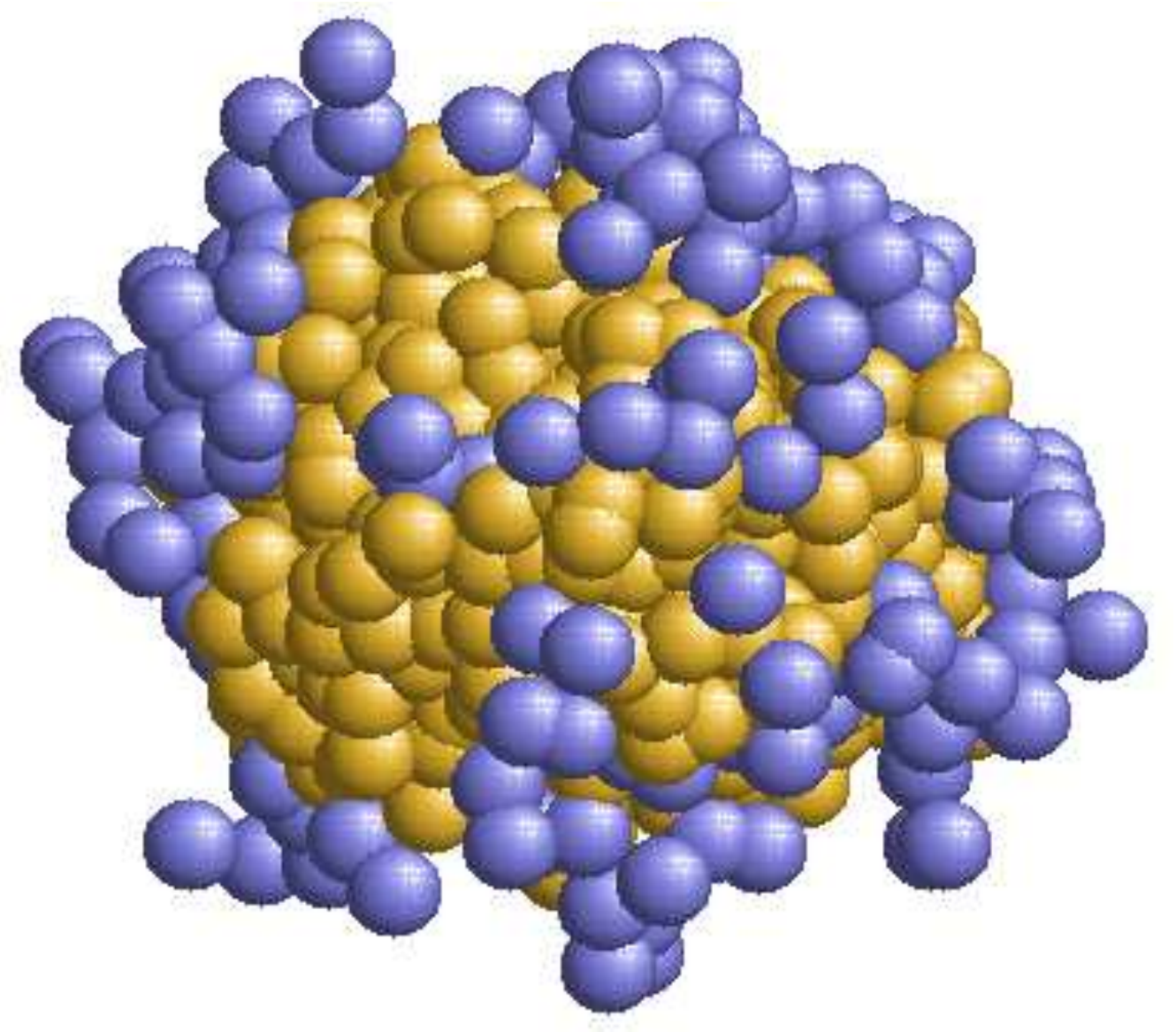}}
\put(20,0){\includegraphics[width=4cm]{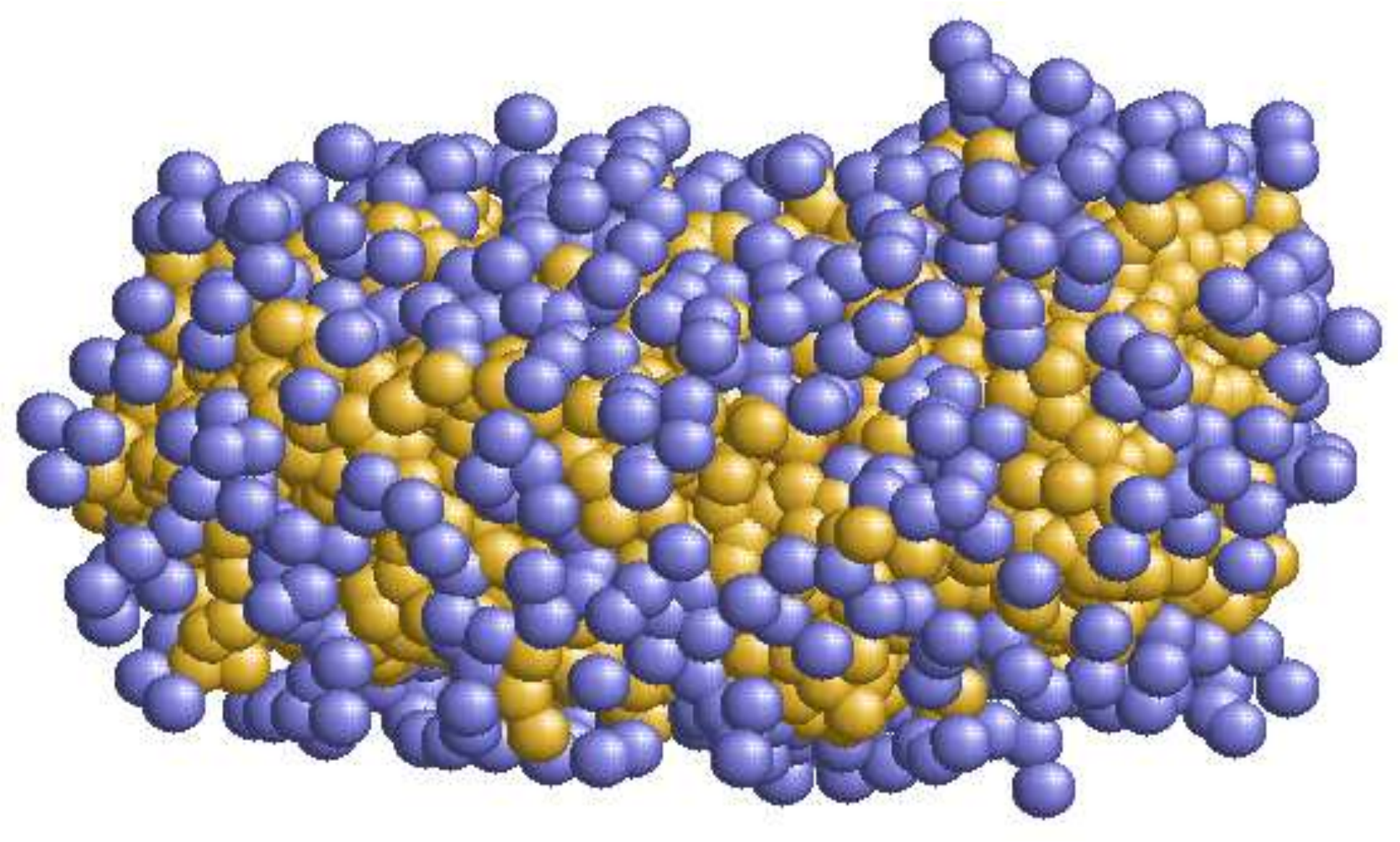}}
\put(170,80){\includegraphics[width=3.5cm]{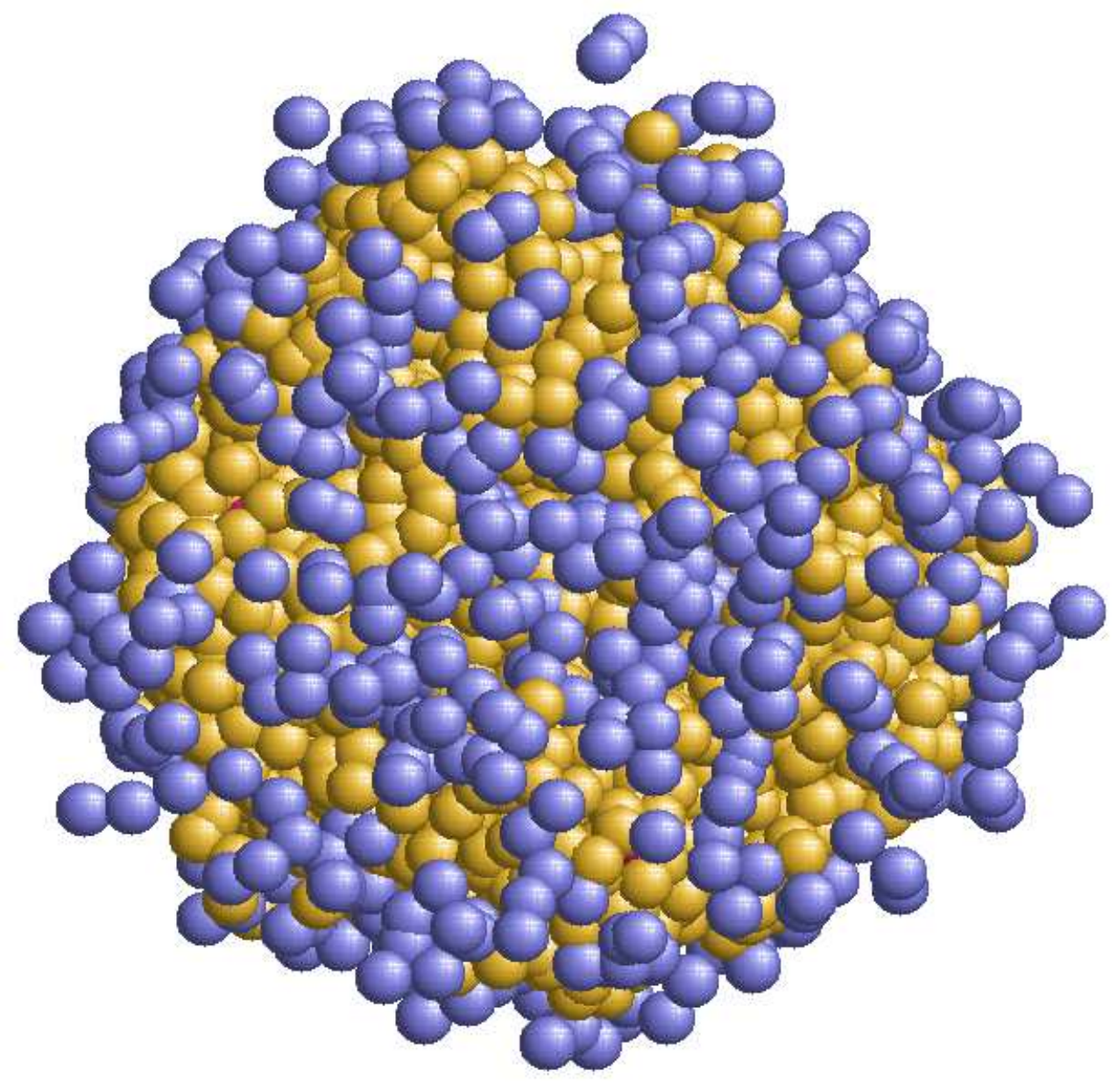}}
\put(170,0){\includegraphics[width=3.5cm]{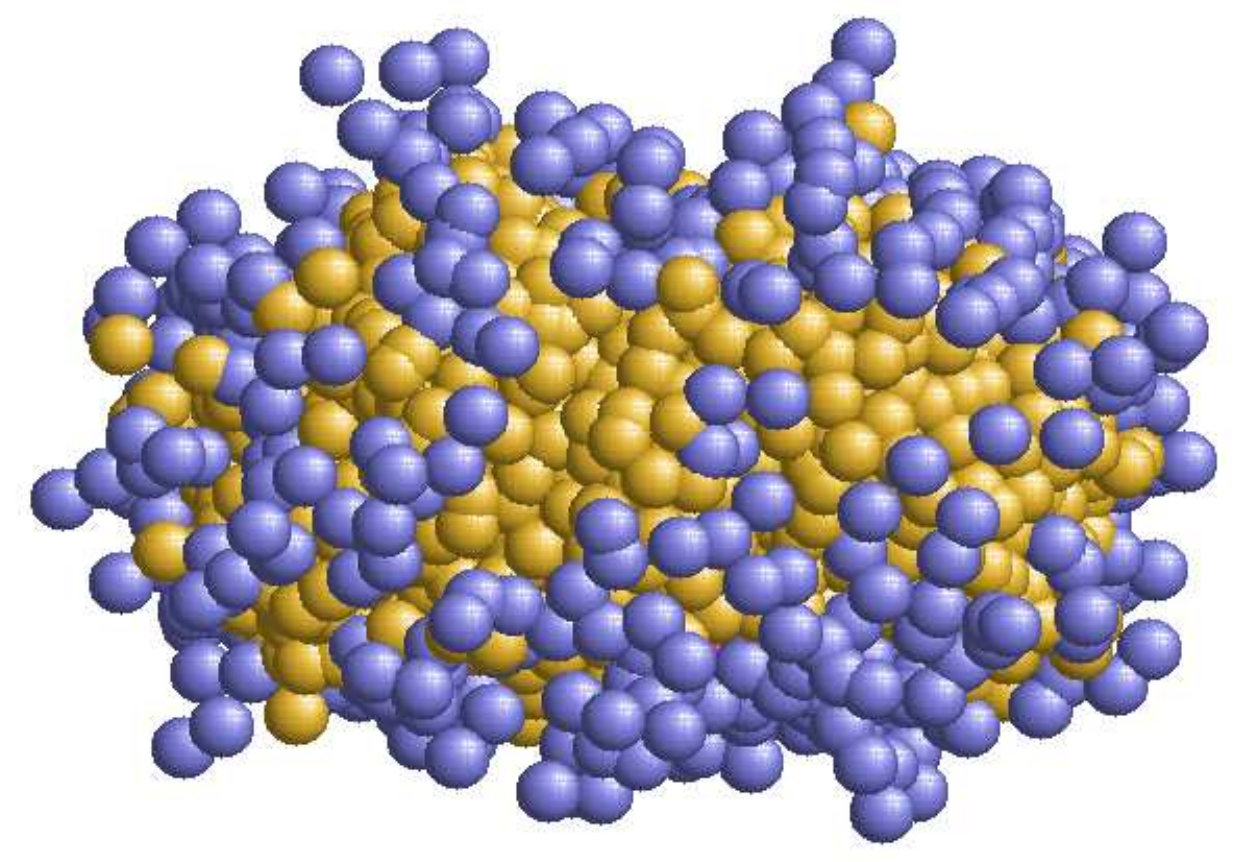}}
\put(300,40){\includegraphics[width=3.5cm]{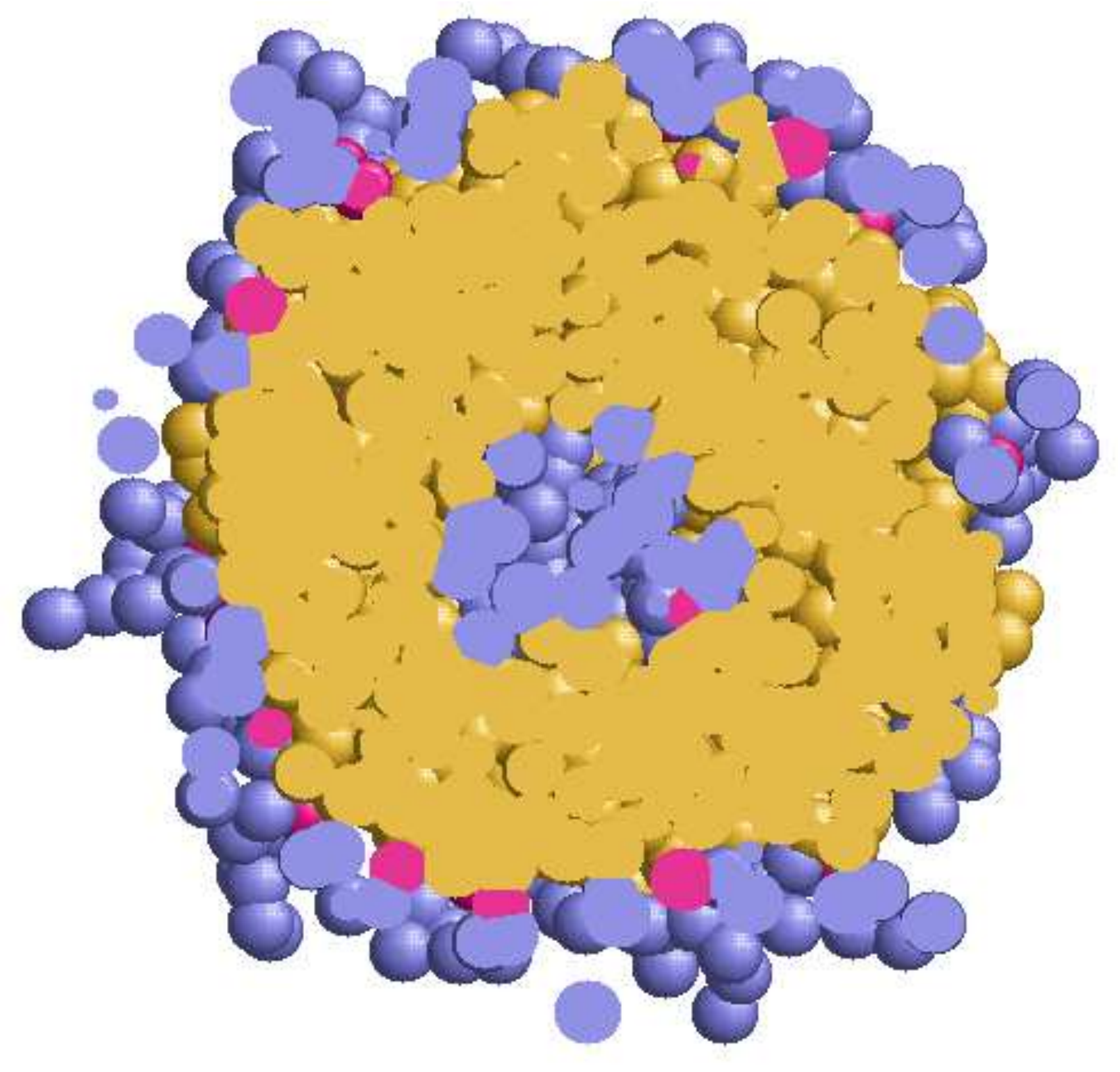}}
\put(45,160){spherical micelle}
\put(40,70){rod-like micelle}
\put(185,185){disc-like micelle}
\put(160,170){top view}
\put(160,70){side view}
\put(310,140){spherical vesicle}
\end{picture}
\caption{\label{snapshots} (Color online) Basic shapes observed for aggregates of amphiphilic stars
(see respective caption next to each frame). Spherical vesicle is shown sliced in the middle
to reveal its internal void, colours of beads follow these in figure~\ref{mol_arch}.}
\end{figure}

On  a classification note, we found four types of aggregate shape: spherical, rod-like and disc-like
micelles, as well as a spherical vesicle, all visualised in figure~\ref{snapshots}. For most molecular
architectures considered here, the increase of $N_{\text a}$ results in the following sequence of shapes:
spherical micelle $\to$ aspherical micelle (rod-like or oscillating between the rod- and disc-like)
$\to$ spherical vesicle.

\begin{figure}[!b]
\vspace{-10mm}
\begin{picture}(0,210)
\put(20,60){\includegraphics[width=2.50cm]{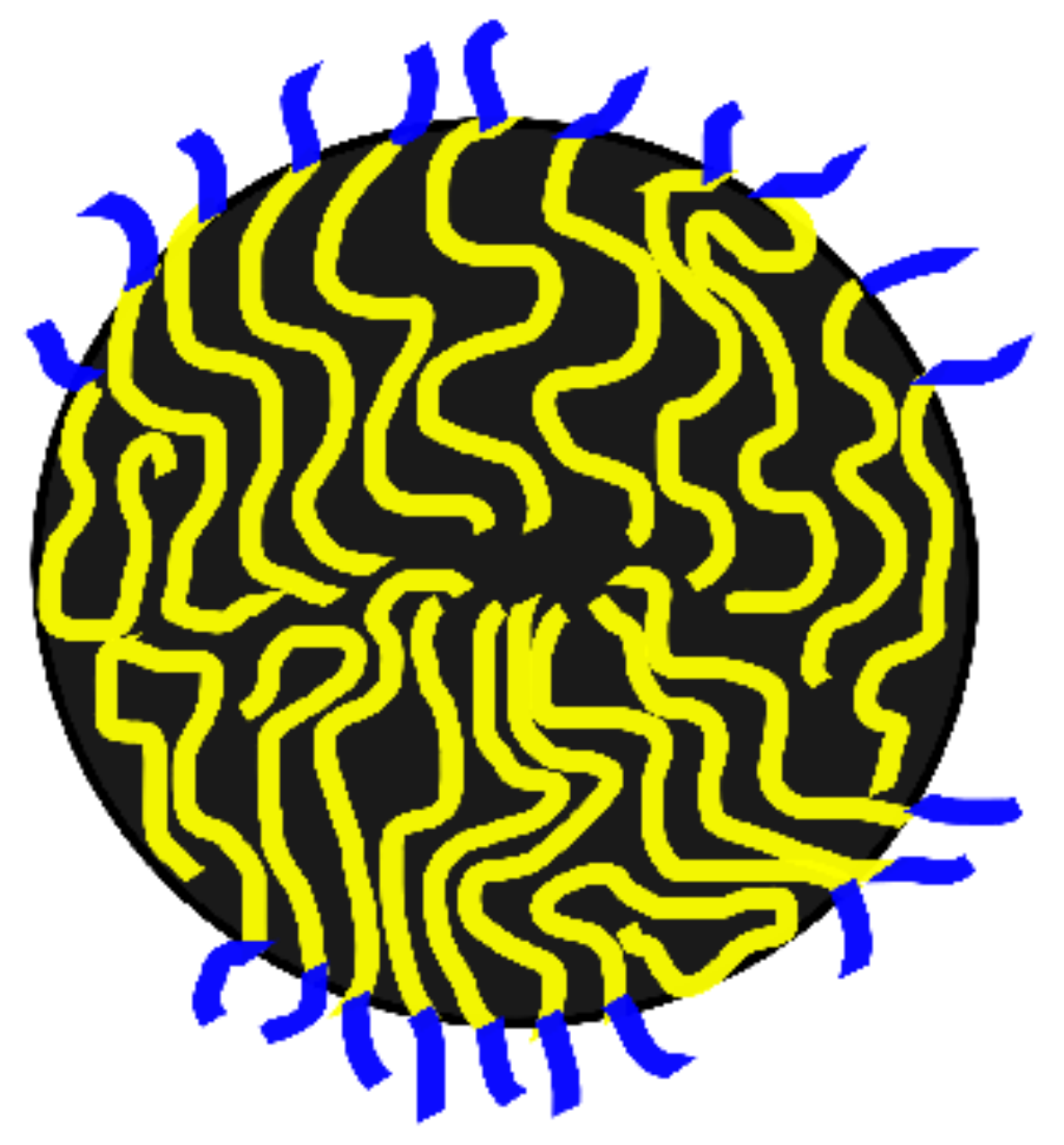}}
\put(130,0){\includegraphics[width=3.8cm]{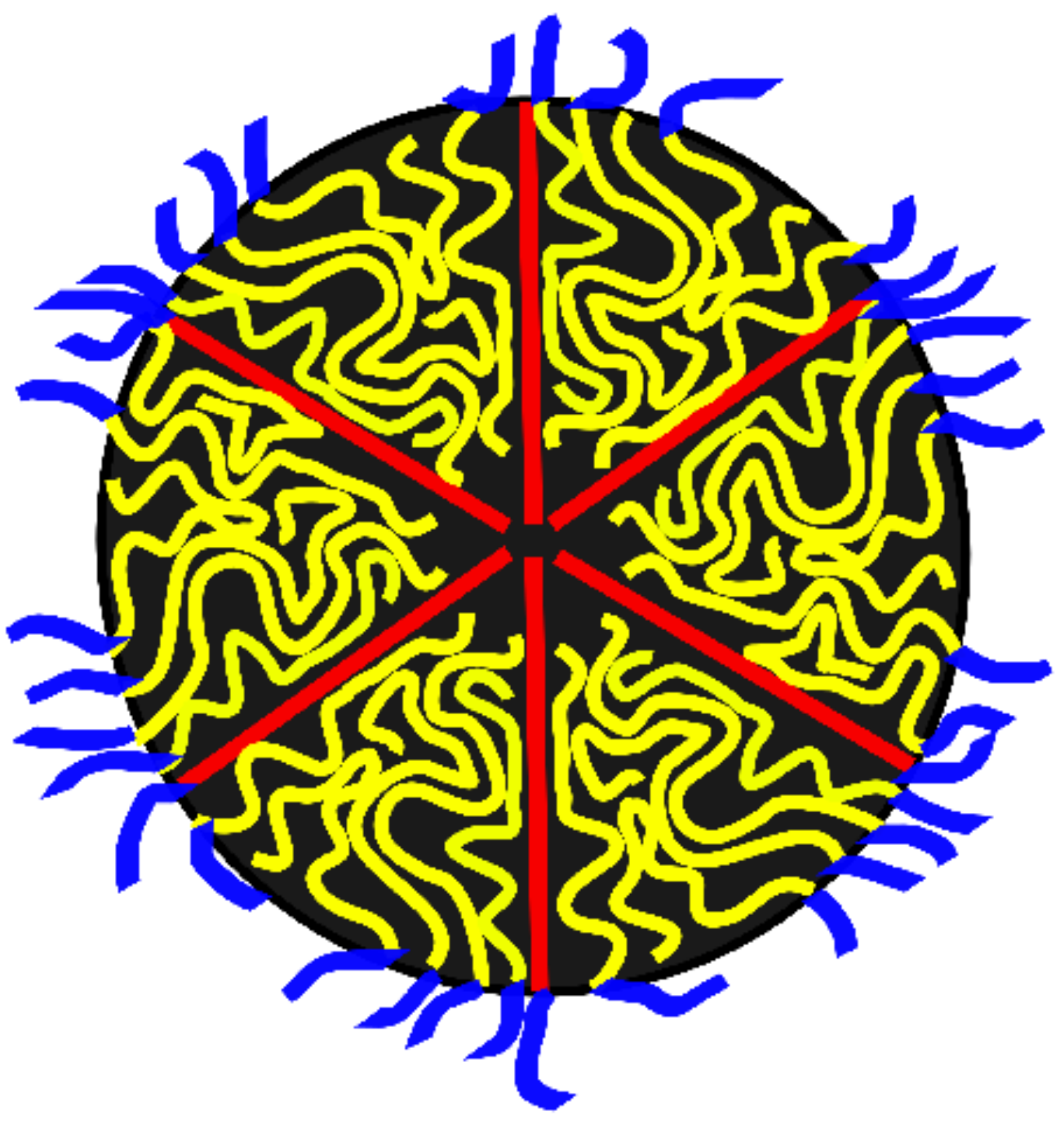}}
\put(125,125){\includegraphics[width=3.8cm]{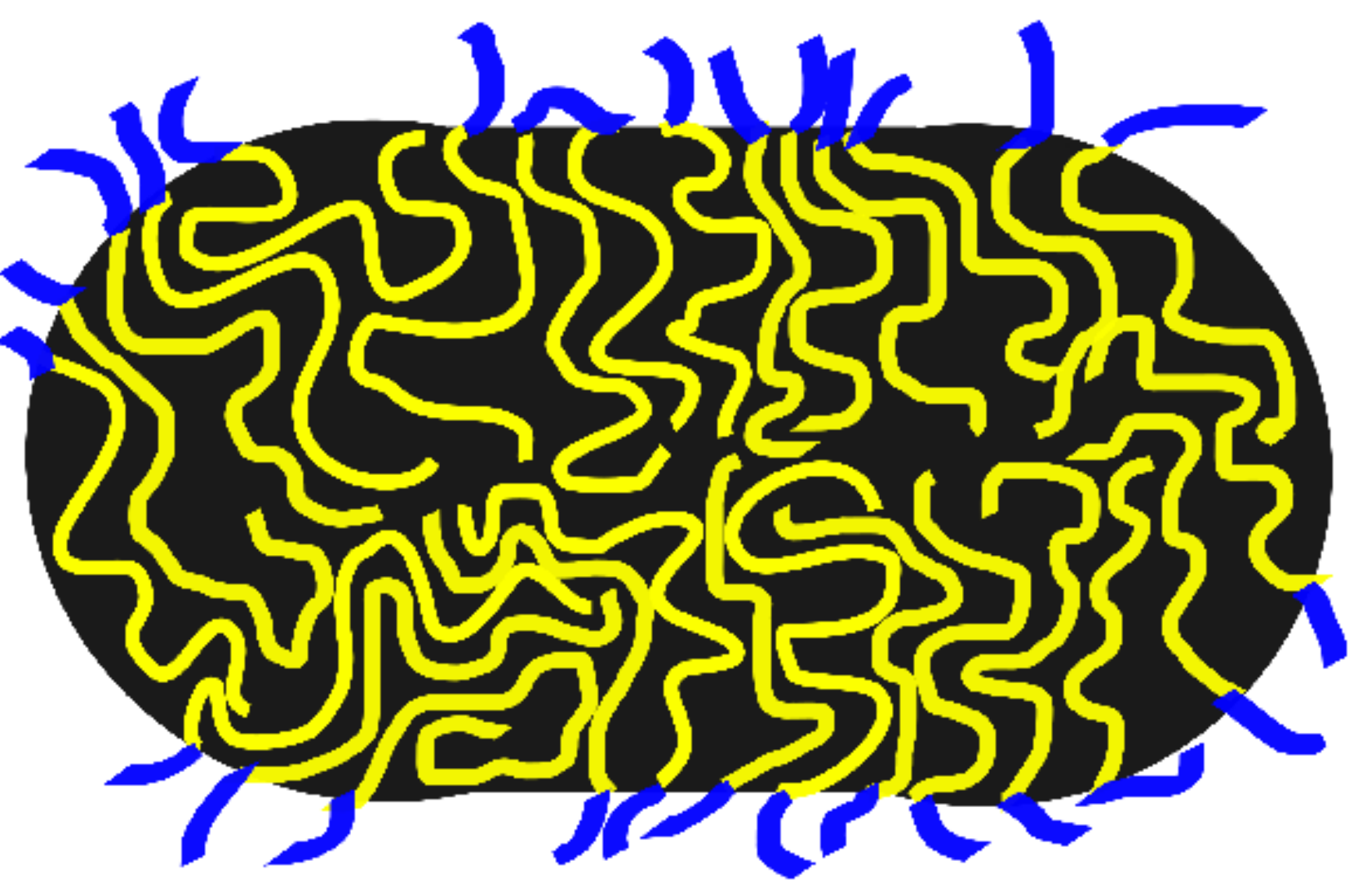}}
\put(280,40){\includegraphics[width=3.8cm]{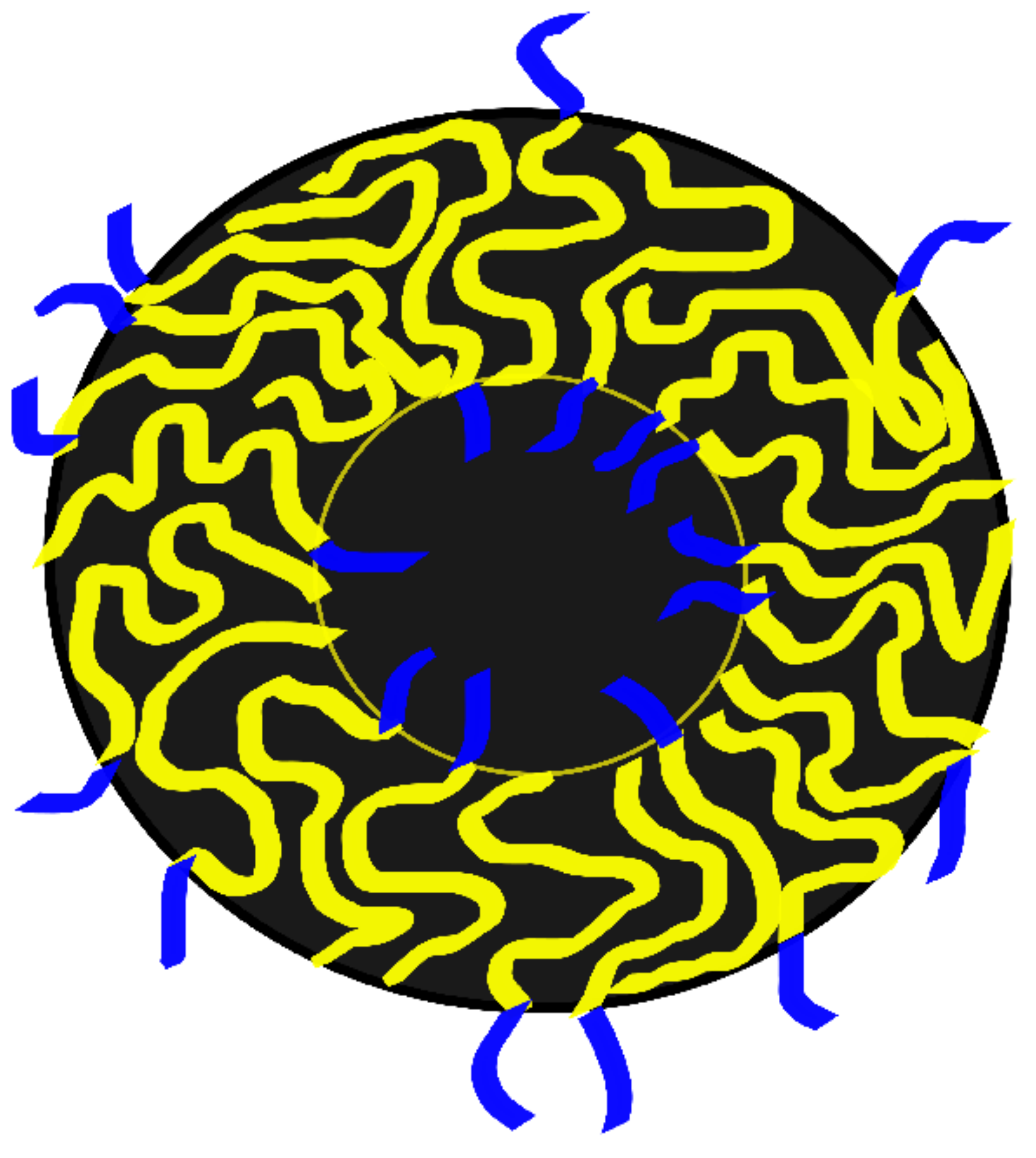}}
\put(55,145){(i)}
\put(130,100){(ii)}
\put(105,180){(iii)}
\put(340,165){(iv)}
\end{picture}
\caption{\label{schematic} (Color online) Schematic representation of the conformations of
polymer chains for the case of (i) spherical micelle, (ii) overgrown spherical micelle, (iii)
rod-like micelle and (iv) spherical vesicle. Colours of beads follow these in figure~\ref{mol_arch}.}
\end{figure}
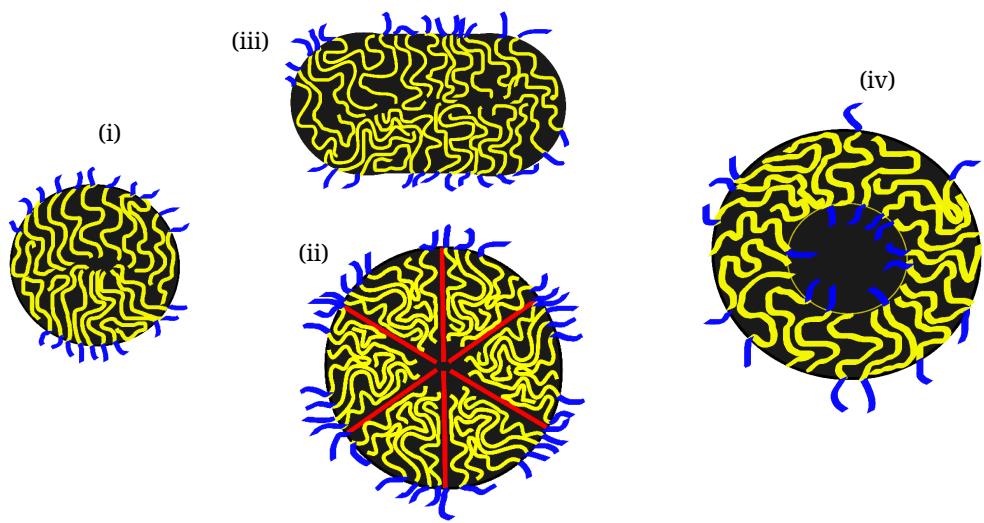

\begin{figure}[!t]
\begin{picture}(0,105)
\put(0,110){\includegraphics[width=4cm,angle=270]{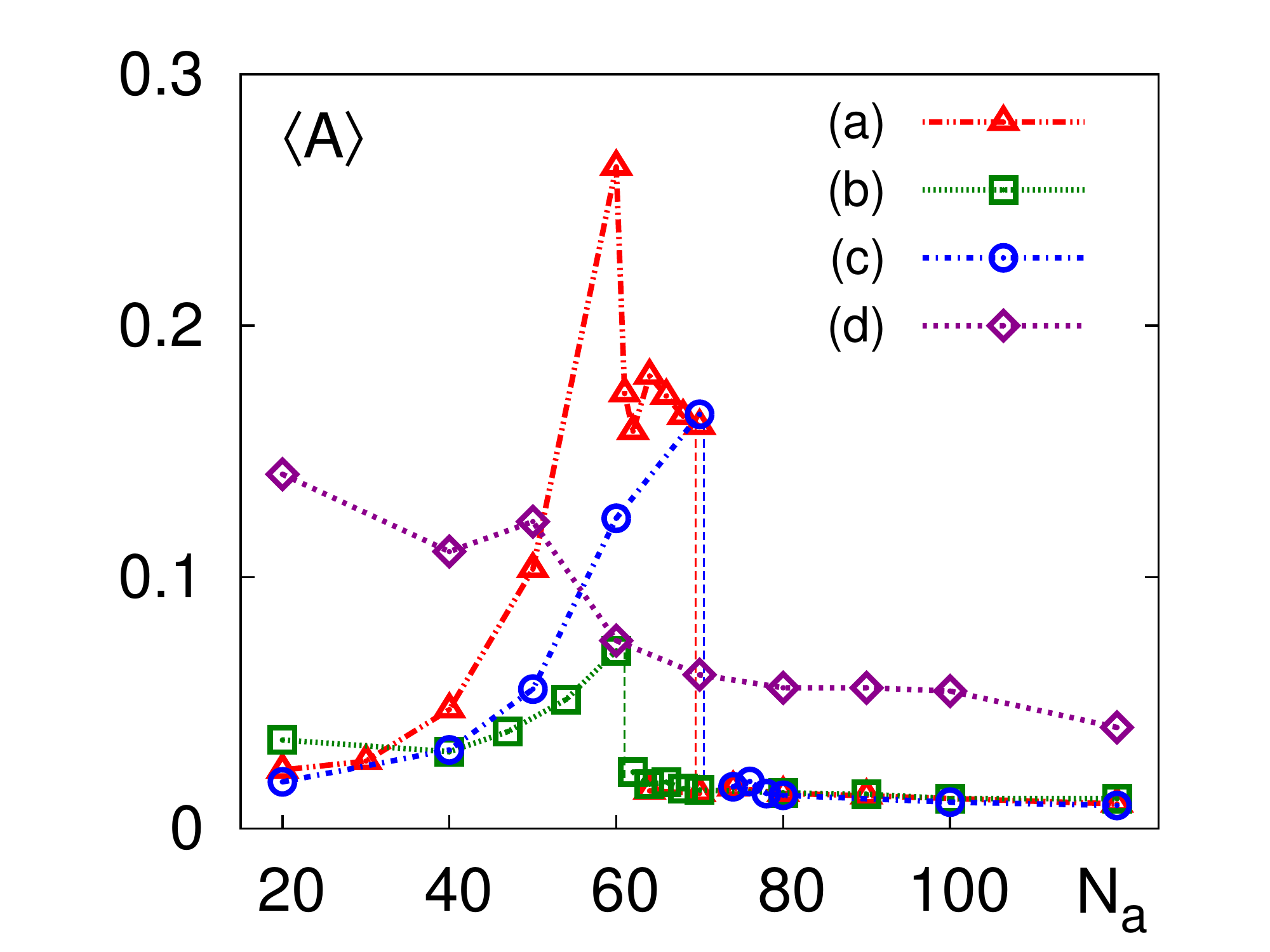}}
\put(130,110){\includegraphics[width=4cm,angle=270]{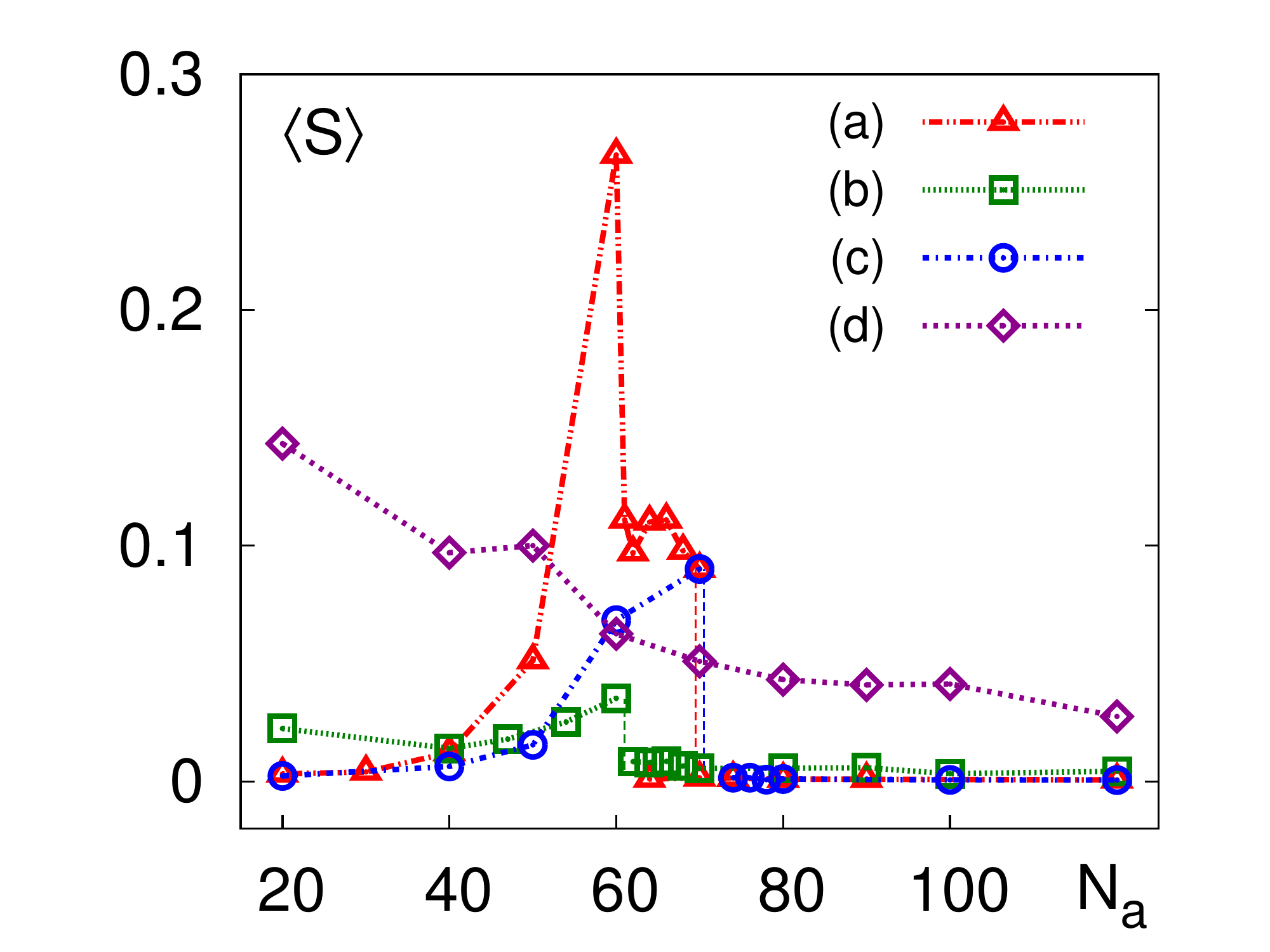}}
\put(260,110){\includegraphics[width=4cm,angle=270]{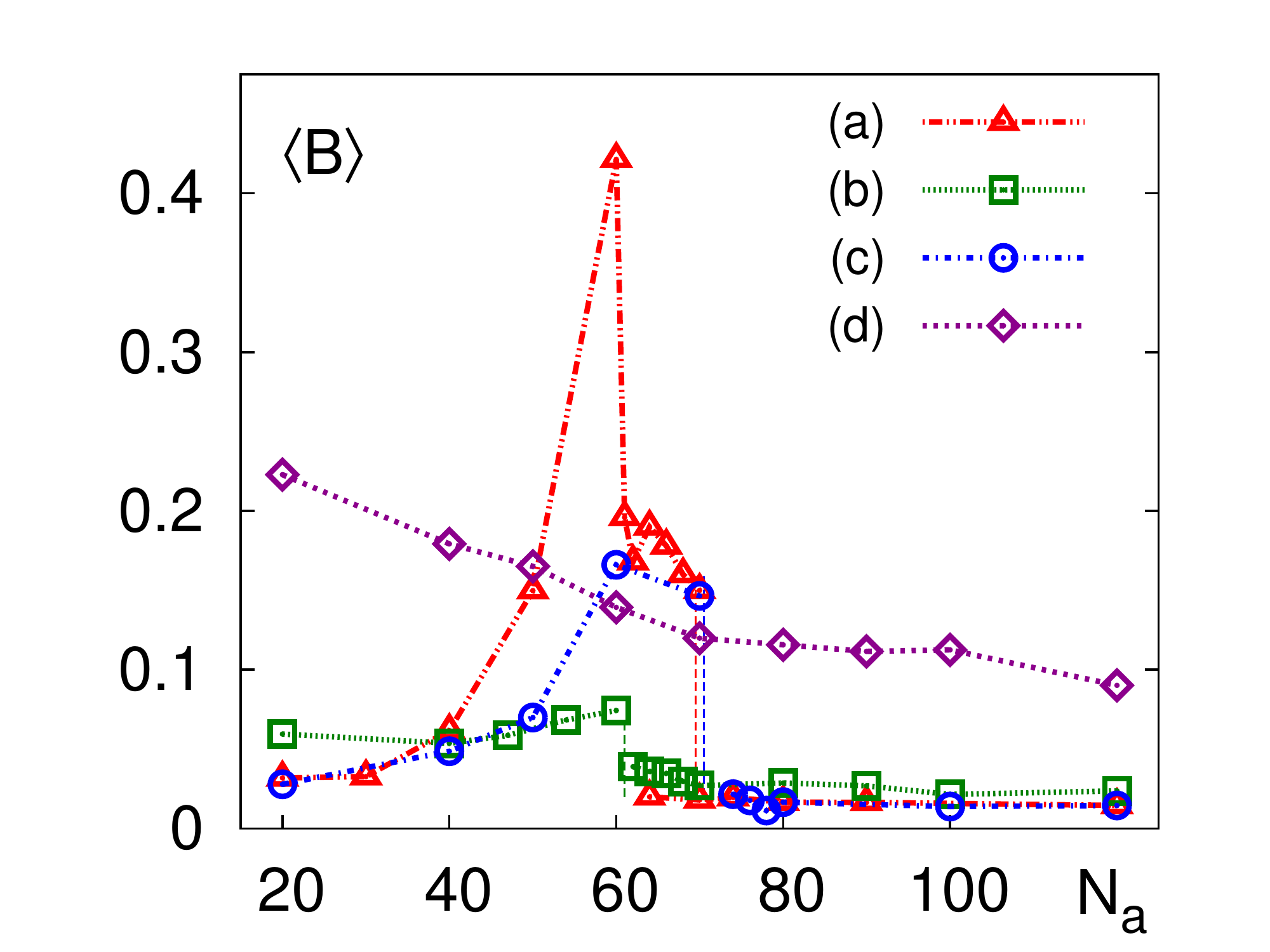}}
\end{picture}
\caption{\label{ASB} (Color online) Comparison of the simple averages for the shape characteristics
$\langle A\rangle$, $\langle S\rangle $ and $\langle B\rangle $ as functions of the
aggregation number $N_{\text a}$. (a)--(d) follow the notations introduced for molecular architectures
in figure~\ref{mol_arch}.}
\end{figure}
This sequence can be understood in terms of the competition between the enthalpic and entropic contributions
to the free energy related to the presence of the aggregate. The enthalpic contribution is minimal when the number
of contacts between the hydrophobic and hydrophilic beads are minimised, which is achieved in the case
of a spherical shape. The entropic contribution is minimised when the hydrophobic chains
within the aggregate display coil-like conformations. Both factors do not compete in the case of a spherical micelle
until its radius $R_{\text a}$ is smaller than the average end-to-end distance $r_{\text e}$ of the hydrophobic chains in a coiled state,
see frame (i) in figure~\ref{schematic}.
With the growth of $N_{\text a}$, there are two options: sperical micelles with $R_{\text a}>r_{\text e}$ shown in frame (ii);
or aspherical micelles shown in frame (iii) of the same figure. For the case (ii), the need to fill-in the center of the
micelles inevitably leads to stretching some of  the hydrophobic chains beyond the $r_{\text e}$ distance
(these chains are shown in red), thus reducing the entropy (i.e., increasing the free energy). The effect
strengthens with the increase of $N_{\text a}$ and, hence, of $R_{\text a}$. Therefore, at a certain value of $N_{\text a}$, the aspherical
aggregate becomes more favourable, when the penalty of aspherical shape is compensated by the gain
from a coiled state for all chains, as shown in frame (iii). A further increase of $N_{\text a}$ forces a higher
asphericity of this shape and a spherical vesicle with the internal void, shown in frame (iv),
becomes more favourable. It combines both overall spherical shape and coiled conformations for the
chains within a hydrophobic shell, which is about $r_{\text a}$ for a single layer vesicle.

Let us consider now how the details of the molecular architecture affect the ``phase
boundaries'' for the shapes shown in figure~\ref{snapshots}. These can be traced from the
dependencies of simple averages for the shape characteristics $\langle A\rangle$,
$\langle S\rangle $ and $\langle B\rangle $ as functions of the aggregation number $N_{\text a}$
shown in figure~\ref{ASB}. For molecular architectures (a)--(c), the behaviour for all three
properties is quite similar. In particular, at small $N_{\text a}\sim 20$, the aggregate is a spherical
micelle. With an increase of $N_{\text a}$, the asphericity grows monotonously until $N_{\text a}<N^*_{\text a}$,
where the $N^*_{\text a}\approx 70$ for the cases (a) and (c) and $N^*_{\text a}\approx 60$ for the case (b).
At $N_{\text a}\sim N^*_{\text a}$, a sharp transition occurs between the spherical micelle ($N_{\text a}<N^*_{\text a}$)
and a spherical vesicle ($N_{\text a}>N^*_{\text a}$), where the shape characteristics drop to zero.
A decline for the shape properties is the smallest for the case (b), whereas for the case (d) no
sharp transition is detected at all and the entire interval $20\leqslant N_{\text a} \leqslant 120$ is characterised
by non-zero values for all three shape characteristics. Their gradual decrease is observed with an
increase of $N_{\text a}$.

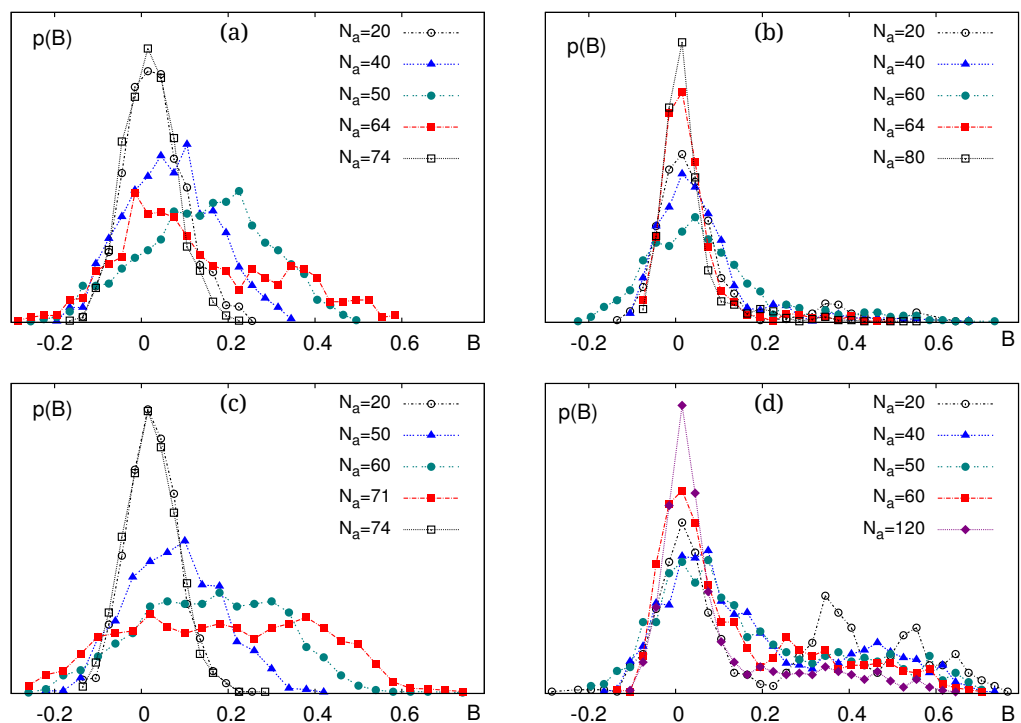
\begin{figure}[!b]
\vspace{-6mm}
\begin{picture}(0,290)
\put(10,280){\includegraphics[width=5cm,angle=270]{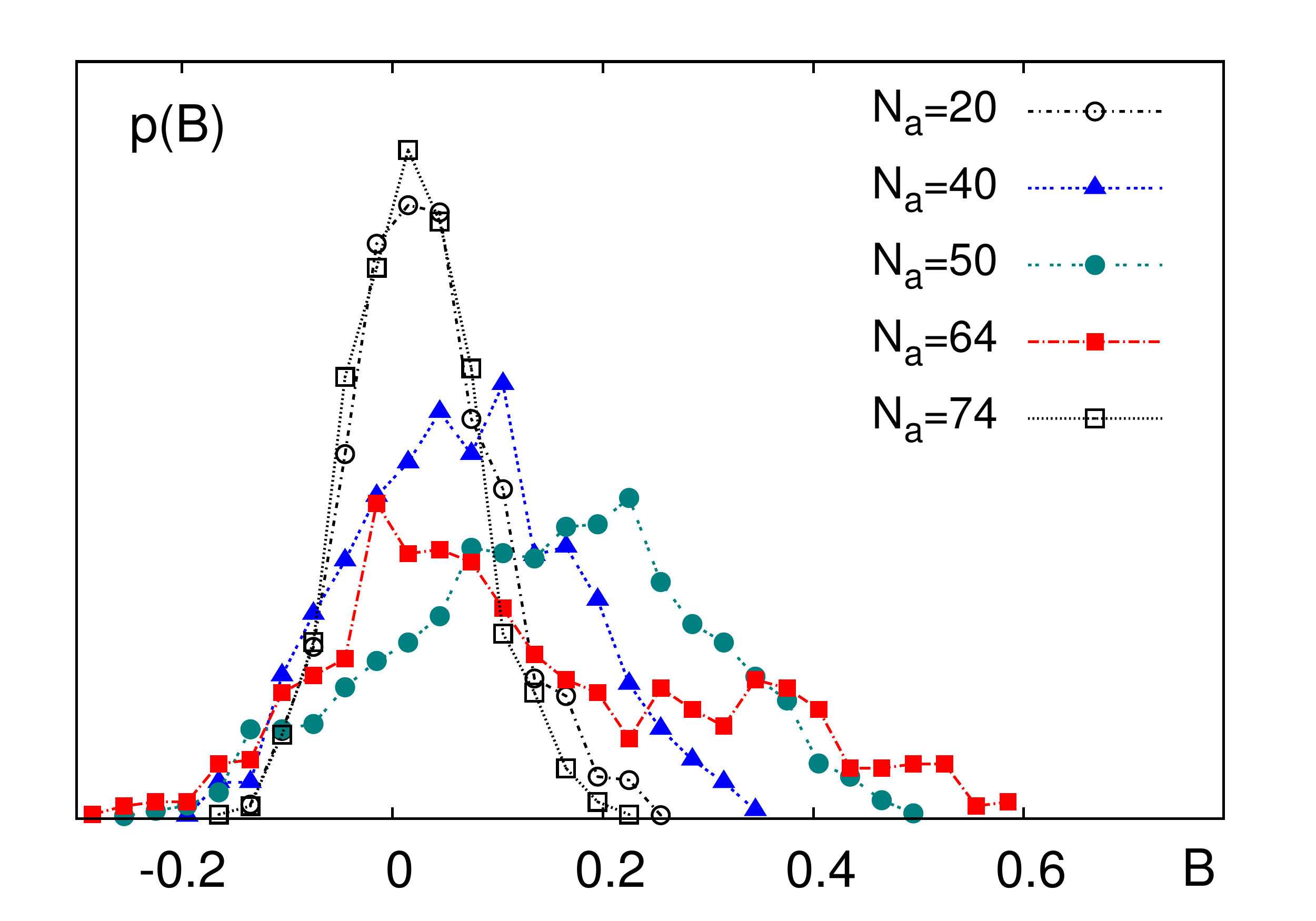}}
\put(210,280){\includegraphics[width=5cm,angle=270]{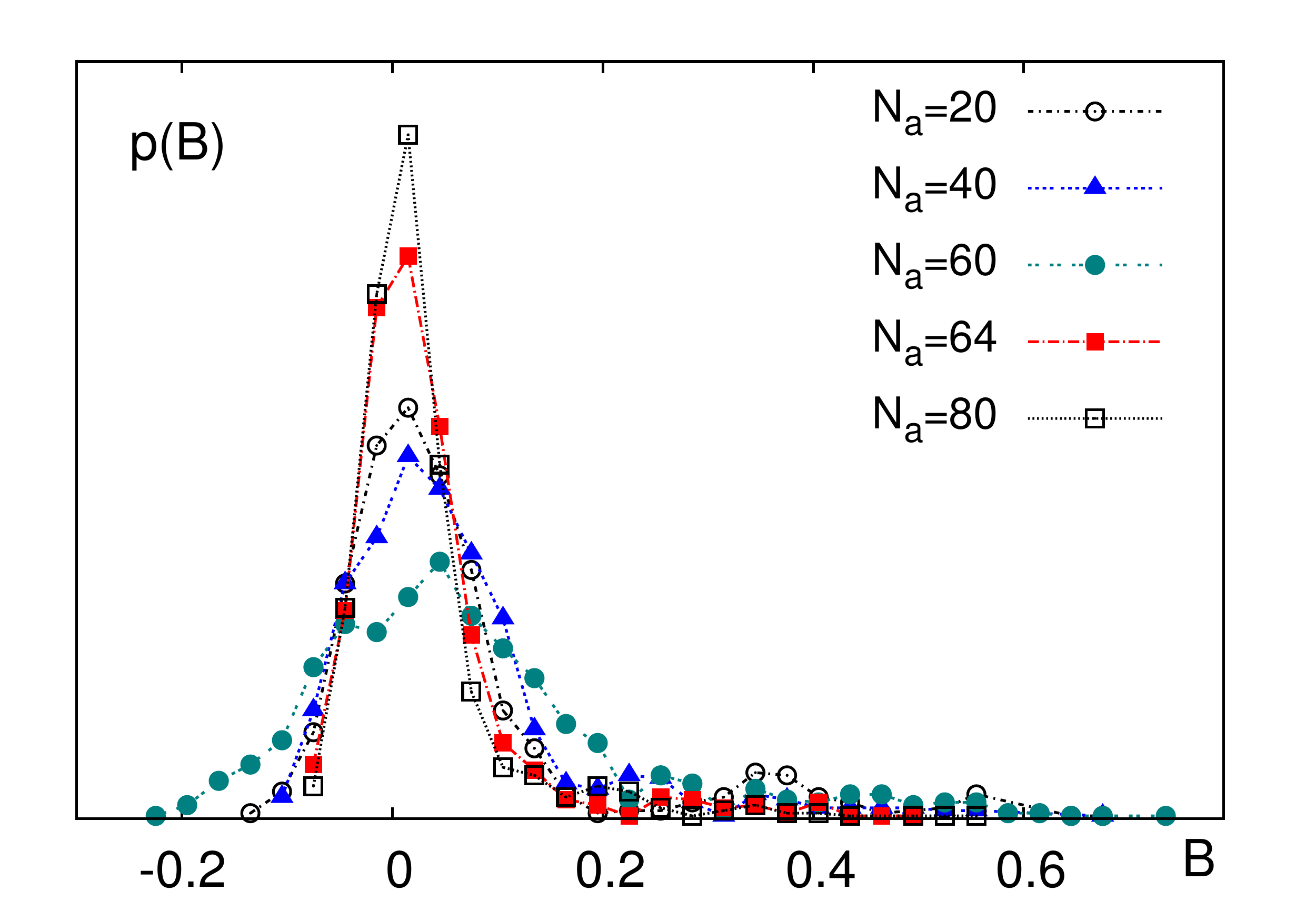}}
\put(10,140){\includegraphics[width=5cm,angle=270]{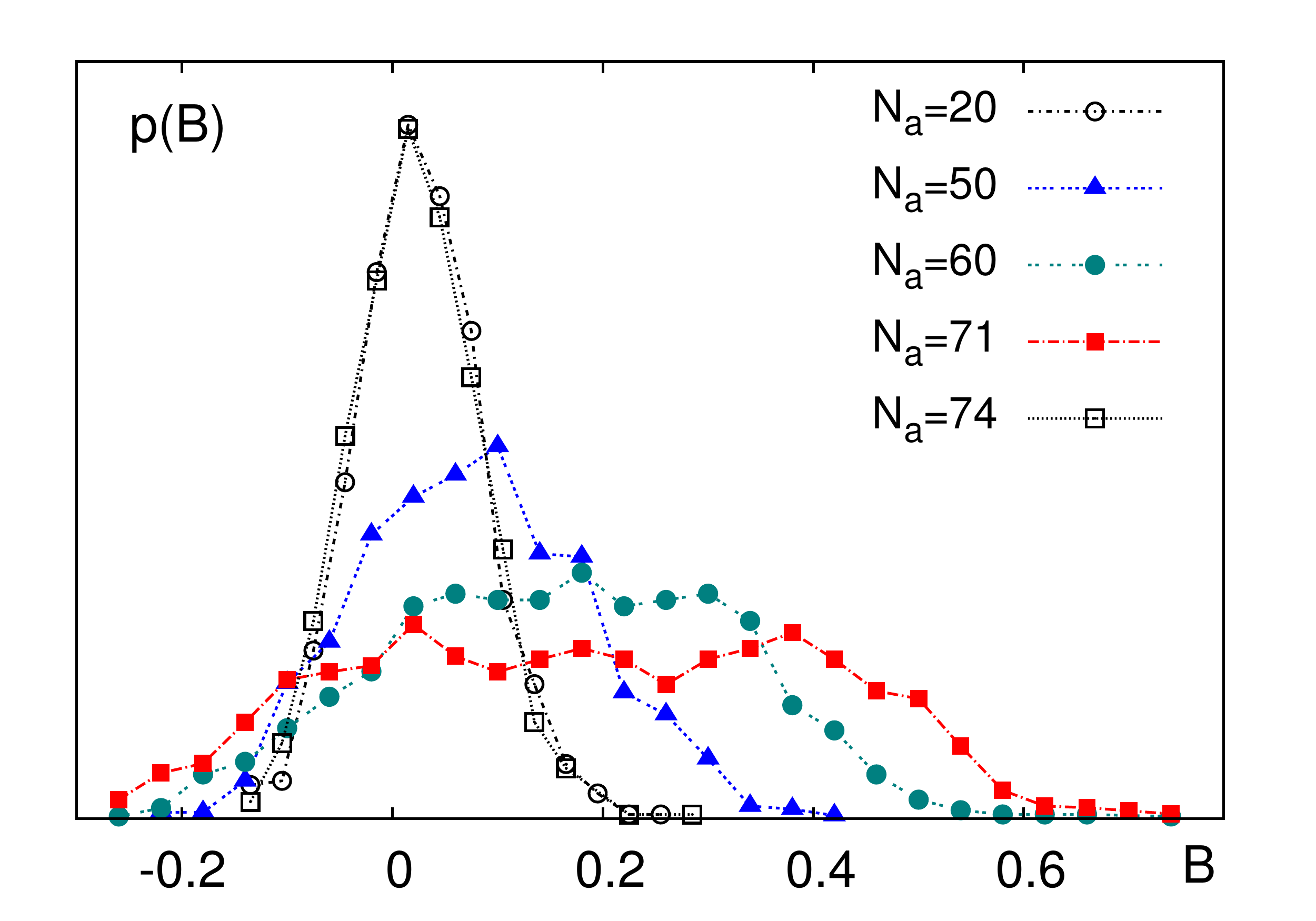}}
\put(210,140){\includegraphics[width=5cm,angle=270]{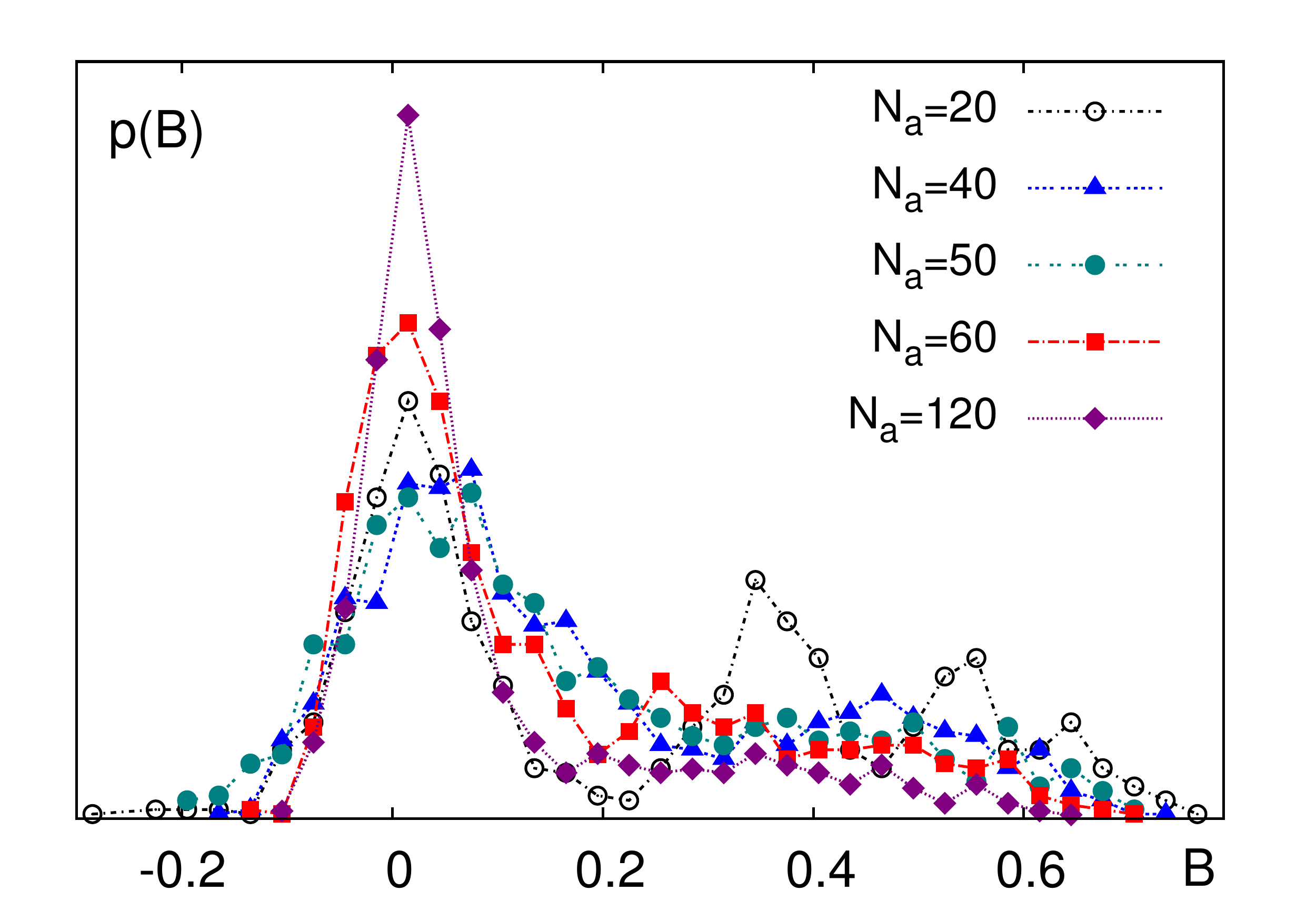}}
\put(100,260){(a)}
\put(300,260){(b)}
\put(100,120){(c)}
\put(300,120){(d)}
\end{picture}
\caption{\label{B_distr} (Color online) Probability distribution $p(B)$ for the shape descriptor $B$
shown for selected values of aggregation number $N_{\text a}$ for the molecular architectures (a)--(d)
listed in figure~\ref{mol_arch}.}
\end{figure}

To shed more light on the behaviour of the shape characteristics, we examine the probability
distribution $p(B)$ for the shape descriptor $B$ in the pretransitional region $N_{\text a}\sim N^*_{\text a}$.
The respective histograms for all four cases (a)--(d) are shown in figure~\ref{B_distr}.
First, let us note that the cases (a) and (c) are very similar showing rather sharp peaks centered
around zero for $N_{\text a}=20$ with the distribution getting broader and shifting towards larger values
of $B$ for larger $N_{\text a}<N^*_{\text a}$. In this interval both disc-like ($B<0$) and rod-like ($B>0$)
shapes are found, albeit the disc-like tendencies are rather weak not reaching the
values below $-0.25$. Above the transition, $N_{\text a}>N^*_{\text a}$, the distribution is back to
a single sharp peak centered around zero indicating a spherical vesicle. Its shape practically overlaps
with the one for the spherical micelle ($N_{\text a}=20$). The case (b) shows similar tendencies
for the probability distribution $p(B)$, but the effect is much weaker. The case (d) is rather different.
At $N_{\text a}=20$, the distribution $p(B)$ shows a series of peaks at relatively large values of $B$
indicating that the spherical shape is not the single favourable one, presumably due to specific
intramolecular constraints of the case (d). The tail of the distribution which spreads towards the
larger values of $B$ persists even for the largest value of $N_{\text a}=120$ being considered.
It lifts the simple average $\langle B \rangle$ to the nonzero values, as seen earlier in figure~\ref{ASB}.

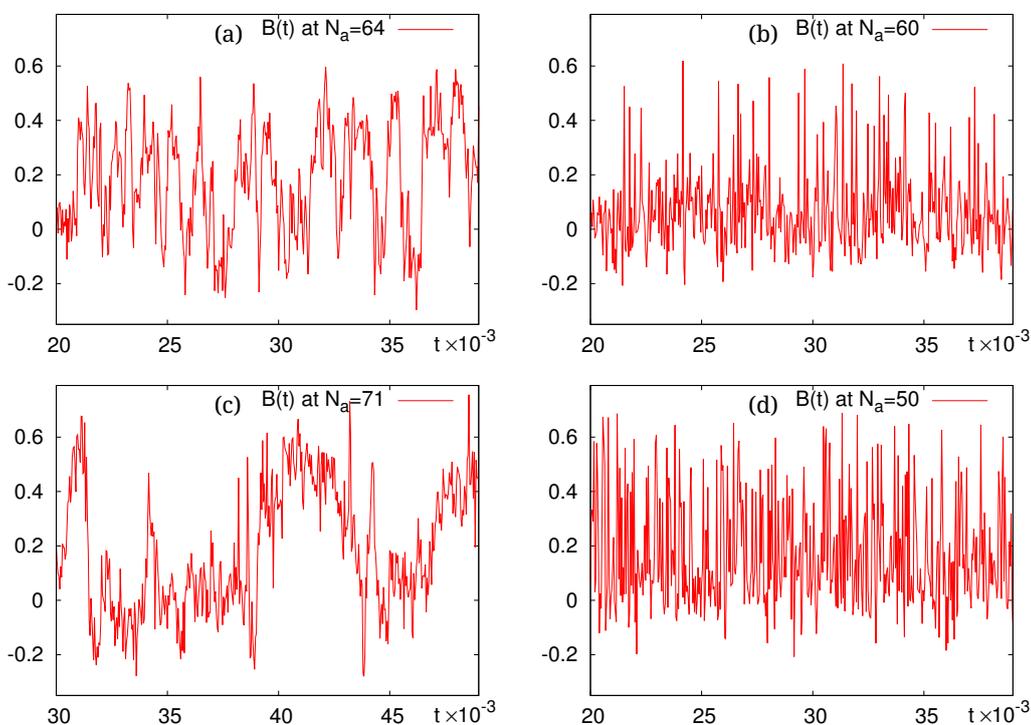
\begin{figure}[!b]
\vspace{-6mm}
\begin{picture}(0,290)
\put(10,280){\includegraphics[width=5cm,angle=270]{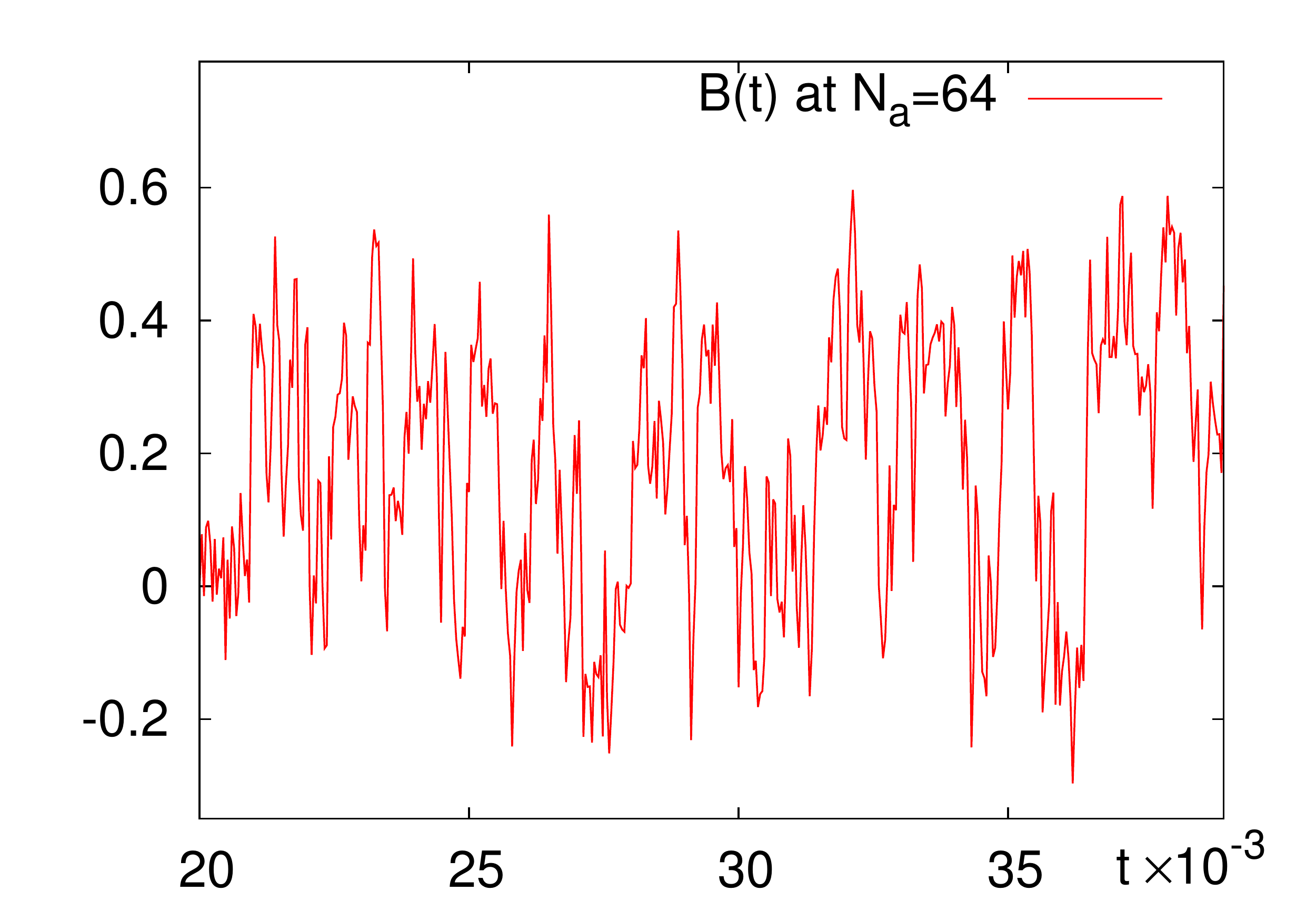}}
\put(210,280){\includegraphics[width=5cm,angle=270]{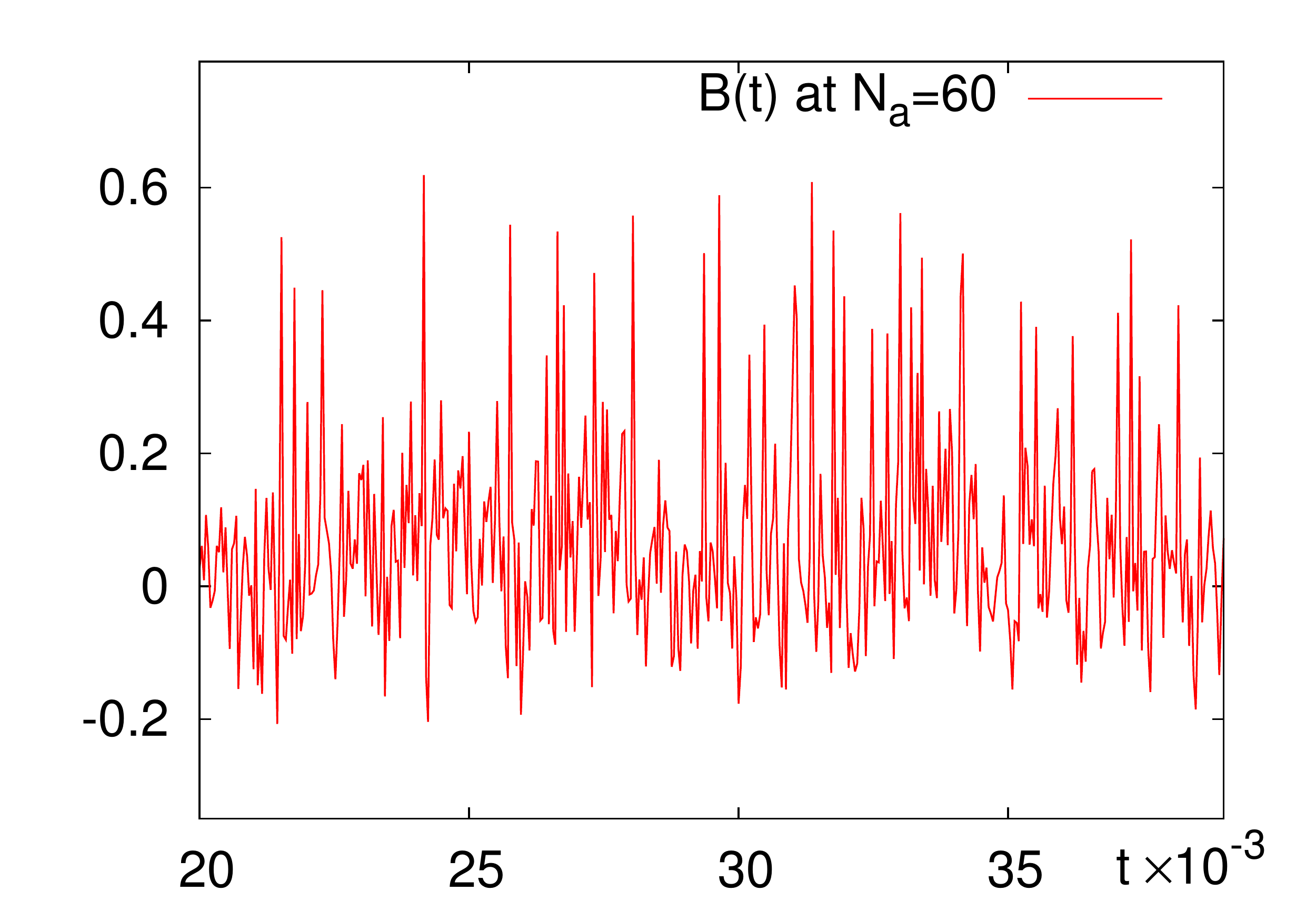}}
\put(10,140){\includegraphics[width=5cm,angle=270]{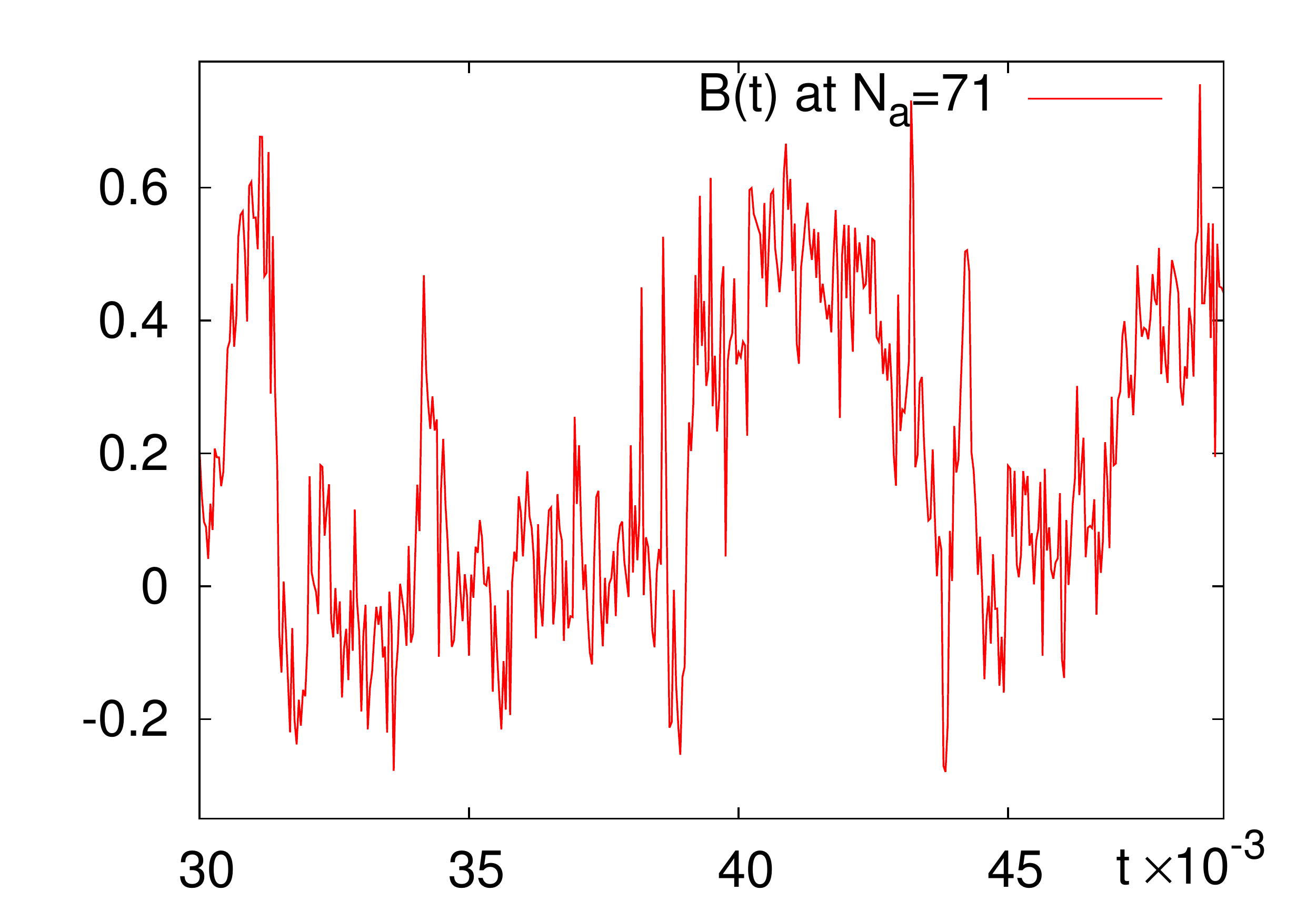}}
\put(210,140){\includegraphics[width=5cm,angle=270]{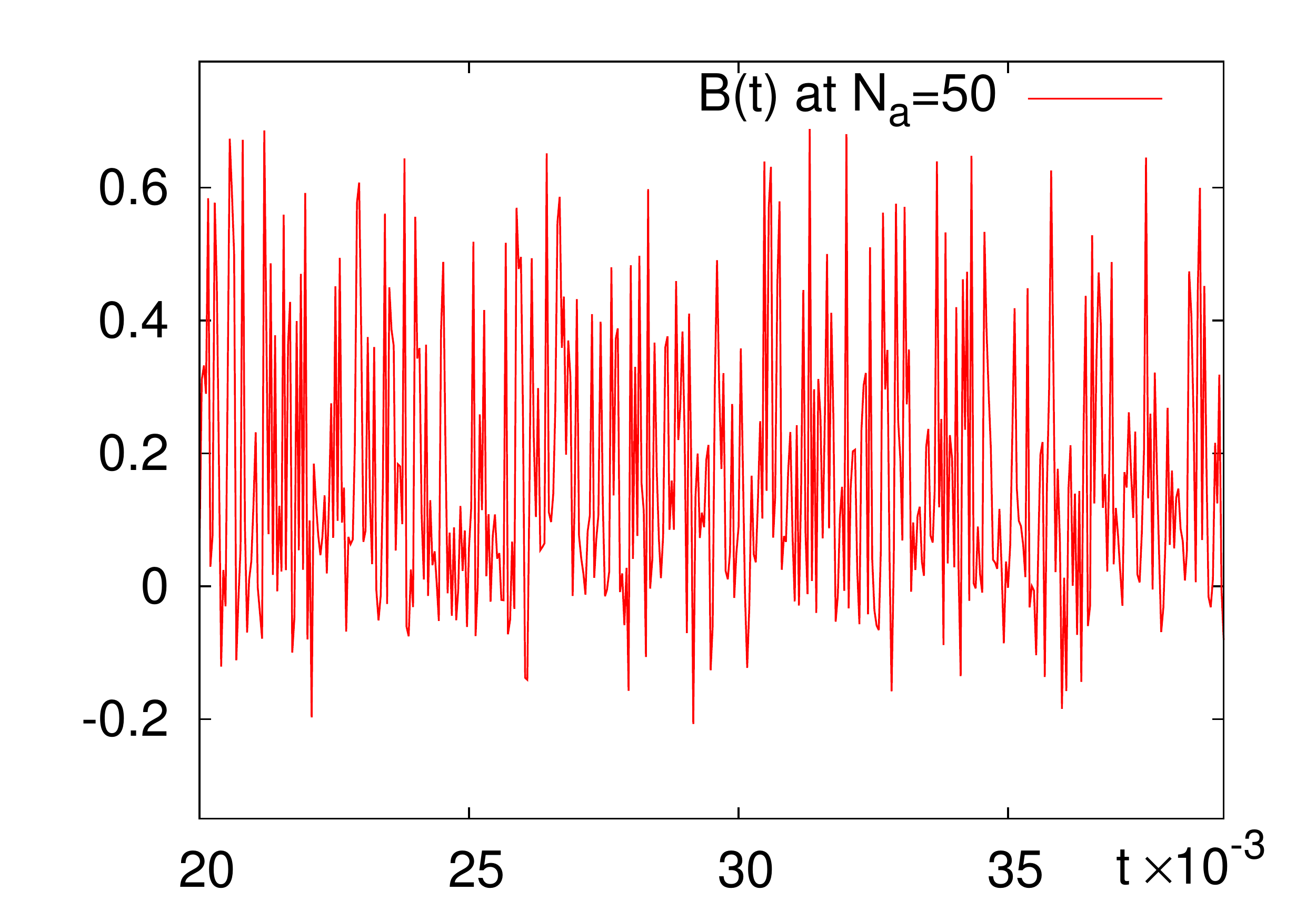}}
\put(100,260){(a)}
\put(300,260){(b)}
\put(100,120){(c)}
\put(300,120){(d)}
\end{picture}
\caption{\label{B_evol} (Color online) Time evolution of the shape descriptor $B$
shown for selected characteristic cases. (a)--(d) stay for the architectures
listed in figure~\ref{mol_arch}.}
\end{figure}

The time evolution for the instant values of the shape descriptor $B(t)$ is shown in figure~\ref{B_evol}
in a pre-transition region for selected characteristic cases (indicated in
each frame). In each case, one can observe a certain level of oscillations of the
aggregate shape occurring between a weak disc-like ($-0.20<B<0$), spherical ($B=0$)
and rod-like ($B>0$) shapes. The period of oscillations is about $2$ time units
for the case~(a) and about $5$ for the case (c). Case (b) is mainly characterised by
values $B\sim 0$, with rare outbursts into $B>0$ region. This
explains that the simple averages for this case shown in figure~\ref{ASB}
are rather small. The case (d) is characterised by fast oscillations between
the weakly disc-like and highly rod-like shapes, resulting in essentially non-zero
simple averages for this case in figure~\ref{ASB}.

One should note that the aspherical shapes of aggregates observed for the amphiphilic stars in this
study, are reminiscent of the worm-like prolate and burger-like oblate multicompartment
micelles observed for the case of the ABC miktoarm stars \cite{Li2004,Cui2006,Xia2006b}.
In the latter case, however, these shapes are stabilised by adding the third component,
whereas for the two-component amphiphilic stars they are prone to large scale oscillations
observed in our work. It is plausible to suggest that these oscillations occur due to a delicate
balance between the enthalpic and entropic contributions, whereas the addition of the hydrophobic drug agent
\cite{Xia2006a} may stabilise the aggregate shape. This will be a subject
of subsequent studies.

%=================================================================
\section{\label{IV}Conclusions}
%=================================================================

The simulation study presented in this work reveals strong dependencies of the
shapes of aggregates formed by amphiphilic stars in water on the details
of their molecular architecture. It has a prospect of the application in the drug delivery
systems, where both the size and the shape of the aggregate is known to play
an important role for the flow of the enveloped agent through the vessels.

Four molecular architectures have been examined: (a) four disjoint linear
diblocks, (b) asymmetric miktoarm polymer, (c) diblock star 1 (hydrophilic parts
pointing outwards) and (c) diblock star 2 (hydrophilic parts next to a central bead).
For all cases, the same general sequence of shapes is found with an increase
of the aggregation number, namely: spherical micelle, aspherical micelle and
a spherical vesicle. The ``phase boundaries''
between these are found to depend on the details of the molecular architecture.
For the case (a)--(c), the transformation between a spherical and aspherical micelle occurs gradually,
whereas the transition from an aspherical micelle into a spherical vesicle is in a form of a sharp transition.
In the case (b), aspherical micelle is less stable and transition to a vesicle occurs at a
lower aggregation number. The case (d) is characterised by gradual transitions
between all the shapes.

Histograms for the probability distributions of the shape descriptor are relatively
narrow for both spherical micelle and spherical vesicle regimes but become wider
next to the micelle-vesicle transition, indicating that a broad range of shapes are possible.
The shape of the aggregate is found to oscillate between the rod-, disc-like and
spherical with the period of oscillations strongly dependent on the molecular
architecture. Both effects of slowing down and acceleration of these oscillations
are found. These findings are relevant for the case of aggregates filled with
water-insoluble drug agent, which will be a topic of the further studies.

%==============================
\section*{Acknowledgements}
%==============================

This work was supported in part by FP7 EU IRSES
projects No. $612707$ ``Dynamics of and in Complex Systems'' and No. $612669$
``Structure and Evolution of Complex Systems with Applications in Physics and Life Sciences'',
and by the Doctoral College for the Statistical Physics of Complex Systems, Leipzig-Lorraine-Lviv-Coventry
$({\mathbb L}^4)$.

\ukrainianpart

\title{Властивості форми агрегатів із амфіфільних зіркових полімерів у воді: дослідження методом дисипативної динаміки}

\author{О.Ю. Калюжний\refaddr{label1,label4},
        Я.М. Ільницький\refaddr{label1,label4}, К. фон Фербер\refaddr{label2,label3,label4}}
\addresses{
\addr{label1} Інститут фізики конденсованих систем НАН України,
вул. Свєнціцького, 1, 79011 Львів, Україна
\addr{label2} Дослідницький центр прикладної математики, Університет Ковентрі, Ковентрі, CV1 5FB, Великобританія
\addr{label3} Дюссельдорфський університет ім. Г. Гайне, D-40225 Дюссельдорф, Німеччина
\addr{label4} Докторський коледж статистичної фізики складних систем, Ляйпціґ-Лотарингія-Львів-Ковентрі
              $({\mathbb L}^4)$, D-04009 Ляйпціґ, Німеччина
}

\makeukrtitle

\begin{abstract}
\tolerance=3000%
В роботі вивчається ефект молекулярної архітектури амфіфільних зіркових
полімерів на форму агрегатів, які вони формують у воді. Полімери та вода
описуються на мезоскопічному рівні, використовуючи метод дисипативної динаміки.
Розглянуті чотири молекулярні архітектури: міктозірки, два типи діблок-зірок
і група лінійних діблок кополімерів, усі з однаковою композицією та молекулярною
масою. Ми розглядаємо початкову конфігурацію у вигляді густого клубка із $N_{\text a}$ молекул,
поміщеного у воду. В рівноважному стані формується агрегат, характеристики форми якого досліджуються
при різних значеннях~$N_{\text a}$. Знайдено чотири форми агрегатів: сферична, стержнеподібна
та дископодібна міцелла та сферична весікула. Оцінені ``фазові границі'' між цими формами
залежно від молекулярної архітектури. У більшості випадків знайдено різкий перехід між
асферичною міцеллою та весікулою. Передперехідна область характеризується
осциляціями властивостей форми із великою амплітудою, частота яких суттєво
залежить від молекулярної архітектури.
\keywords зіркові полімери, амфіфіли, міцелли, весікули, дисипативна динаміка

\end{abstract}

\end{document}